\documentclass[reqno]{amsart}

\usepackage{mathtools,amssymb,physics}
\mathtoolsset{showonlyrefs,showmanualtags}
\usepackage[thinc]{esdiff}
\usepackage{graphicx}
\usepackage{xcolor}
\usepackage{float}
\usepackage[hidelinks]{hyperref}
\usepackage[nameinlink]{cleveref}
\usepackage[T1]{fontenc} 
\usepackage{lmodern} 
\usepackage[utf8]{inputenc} 
\usepackage{enumitem}
\setlist[enumerate,1]{label={(\roman*)}}
\usepackage[left=2cm, right=2cm, top=3cm, bottom=3cm]{geometry}
\usepackage{setspace}
\usepackage[
    backend=biber,
    style=alphabetic,
    maxnames=4,
    maxalphanames=4,
    date=year,
    doi=true,
    url=false,
    isbn=false,
    eprint=true
]{biblatex}
\AtEveryBibitem{%
  \clearfield{url}%
  \clearfield{urldate}%
  \clearfield{pubstate}%
  \clearfield{issn}%
  \clearfield{isbn}%
  \clearfield{eprintclass}%
  \clearlist{language}%
}
\graphicspath{ {images/} }

\usepackage{titlesec}
\titleformat{\chapter}
{\normalfont\Large\bfseries}{\thechapter}{1em}{}
\titleformat{\section}
{\normalfont\normalsize\bfseries}{\thesection}{1em}{}
\titleformat{\subsection}
{\normalfont\normalsize\bfseries}{\thesubsection}{1em}{}
\titleformat{\subsubsection}
{\normalfont\normalsize\bfseries}{\thesubsubsection}{1em}{}
\titleformat{\paragraph}[runin]
{\normalfont\normalsize\bfseries}{\theparagraph}{1em}{}
\titleformat{\subparagraph}[runin]
{\normalfont\normalsize\bfseries}{\thesubparagraph}{1em}{}

\usepackage{amsthm}

\usepackage{aliascnt}

\theoremstyle{plain}
\newtheorem{theorem}{Theorem}
\newtheorem*{theorem*}{Theorem}
\numberwithin{theorem}{section}

\newaliascnt{corollary}{theorem}
\newtheorem{corollary}[corollary]{Corollary}
\aliascntresetthe{corollary}
\numberwithin{corollary}{section}

\newaliascnt{lemma}{theorem}
\newtheorem{lemma}[lemma]{Lemma}
\aliascntresetthe{lemma}
\numberwithin{lemma}{section}

\newaliascnt{proposition}{theorem}
\newtheorem{proposition}[proposition]{Proposition}
\aliascntresetthe{proposition}
\numberwithin{proposition}{section}

\newaliascnt{conjecture}{theorem}
\newtheorem{conjecture}[conjecture]{Conjecture}
\aliascntresetthe{conjecture}
\numberwithin{conjecture}{section}

\theoremstyle{definition}

\newaliascnt{definition}{theorem}
\newtheorem{definition}[definition]{Definition}
\aliascntresetthe{definition}
\numberwithin{definition}{section}

\newaliascnt{example}{theorem}
\newtheorem{example}[example]{Example}
\aliascntresetthe{example}
\numberwithin{example}{section}

\newaliascnt{remark}{theorem}
\newtheorem{remark}[remark]{Remark}
\aliascntresetthe{remark}
\numberwithin{remark}{section}

\newaliascnt{question}{theorem}
\newtheorem{question}[question]{Question}
\aliascntresetthe{question}
\numberwithin{question}{section}

\let\originalleft\left
\let\originalright\right
\renewcommand{\left}{\mathopen{}\mathclose\bgroup\originalleft}
\renewcommand{\right}{\aftergroup\egroup\originalright}

\newcommand{\cA}{\mathcal{A}}
\newcommand{\cB}{\mathcal{B}}
\newcommand{\cC}{\mathcal{C}}

\newcommand{\cE}{\mathcal{E}}
\newcommand{\cF}{\mathcal{F}}

\newcommand{\cH}{\mathcal{H}}

\newcommand{\cK}{\mathcal{K}}

\newcommand{\cR}{\mathcal{R}}
\newcommand{\cS}{\mathcal{S}}

\newcommand{\cV}{\mathcal{V}}

\newcommand{\cX}{\mathcal{X}}
\newcommand{\cY}{\mathcal{Y}}

\newcommand{\bC}{\mathbb{C}}
\newcommand{\bF}{\mathbb{F}}

\newcommand{\bR}{\mathbb{R}}

\newcommand{\bU}{\mathbb{U}}
\newcommand{\bZ}{\mathbb{Z}}

\newcommand{\rbr}[1]{\left( #1 \right)}
\newcommand{\sqbr}[1]{\left[ #1 \right]}
\newcommand{\crbr}[1]{\left\{ #1 \right\}}

\newcommand{\kket}[1]{\ket{\!\ket{#1}\!}}
\newcommand{\bbra}[1]{\bra{\!\bra{#1}\!}}
\newcommand{\kketbbra}[2]{\kket{#1}\bbra{#2}}
\newcommand{\bbrakket}[2]{\bbra{#1}\kket{#2}}

\newcommand{\bo}[1]{\cB\left(#1\right)}
\newcommand{\bu}[1]{\bU\left(#1\right)}

\newcommand{\bsa}[1]{\cB\left(#1\right)_{\mathrm{sa}}}

\newcommand{\hoh}{\cH\otimes\cH^*}

\DeclareMathOperator{\ad}{ad}
\DeclareMathOperator{\Ad}{Ad}
\DeclareMathOperator{\Cost}{Cost}

\DeclareMathOperator{\id}{id}

\DeclareMathOperator{\Iso}{Iso}
\DeclareMathOperator{\Lin}{Lin}

\DeclareMathOperator{\ran}{ran}
\renewcommand{\Re}{Re}
\renewcommand{\span}{\mathrm{span}}

\DeclareMathOperator{\ST}{ST}
\DeclareMathOperator{\SU}{SU}

\DeclareMathOperator{\Var}{Var}

\newcommand{\be}{\begin{equation}\begin{aligned}}
\newcommand{\ee}{\end{aligned}\end{equation}}
\newcommand{\bp}{\begin{proposition}}
\newcommand{\ep}{\end{proposition}}
\newcommand{\bpr}{\begin{proof}}
\newcommand{\epr}{\end{proof}}
\newcommand{\bc}{\begin{corollary}}
\newcommand{\ec}{\end{corollary}}
\newcommand{\bde}{\begin{definition}}
\newcommand{\ede}{\end{definition}}
\newcommand{\bco}{\begin{conjecture}}
\newcommand{\eco}{\end{conjecture}}
\newcommand{\bex}{\begin{example}}
\newcommand{\eex}{\end{example}}
\newcommand{\bl}{\begin{lemma}}
\newcommand{\el}{\end{lemma}}
\newcommand{\br}{\begin{remark}}
\newcommand{\er}{\end{remark}}
\newcommand{\bi}{\begin{enumerate}}
\newcommand{\ei}{\end{enumerate}}
\newcommand{\ii}{\item}
\newcommand{\bq}{\begin{question}}
\newcommand{\eq}{\end{question}}
\newcommand{\bt}{\begin{theorem}}
\newcommand{\et}{\end{theorem}}
\newcommand{\btu}{\begin{theorem*}}
\newcommand{\etu}{\end{theorem*}}
\newtheorem*{maintheorempreview}{%
  \protect\hyperref[thm:main]{%
    Theorem~\protect\ref*{thm:main}%
  }%
}

\author[Gergely Bunth]{Gergely Bunth}
\address{Gergely Bunth, HUN-REN Alfréd Rényi Institute of Mathematics\\ Reáltanoda u. 13-15.\\Budapest H-1053\\ Hungary\\ and Department of Analysis and Operations Research, Institute of Mathematics, Budapest University of Technology and Economics\\ Műegyetem rkp. 3. \\ Budapest H-1111 \\ Hungary}

\email{bunth.gergely@renyi.hu}

\begin{document}

\subjclass[2020]{Primary: 49Q22; 81P16. Secondary: 81R05; 15A63}

\keywords{quantum optimal transport, symmetric quantum Wasserstein distances, Wigner-type isometries, isotropic quadratic costs, Hilbert--Schmidt tight frames}

\thanks{Bunth was supported by the Momentum program of the Hungarian Academy of Sciences under grant agreement no. LP2021-15/2021, and by the Hungarian National Research, Development and Innovation Office (NKFIH) under grant agreement no. Excellence\_151232.}

\title[Wigner symmetries single out symmetric Wasserstein distances] {Wigner symmetries single out symmetric Wasserstein distances in all finite dimensions}

\begin{abstract}

We study the quantum Wasserstein distances of De Palma and Trevisan associated with quadratic costs generated by self-adjoint observables in arbitrary finite dimension. We first show that any positive semidefinite cost operator is determined by the corresponding distance on pairs of pure states. For nonzero observable-generated quadratic costs, we then prove that the full isometry monoid consists exactly of unitary and antiunitary conjugations if and only if the distance is invariant under unitary conjugations on pure states. These conditions are equivalent to isotropy of the cost on the traceless subspace. They are also equivalent to a geometric characterization: two states attain the Wasserstein diameter exactly when their supports are orthogonal. It already suffices to impose this condition on pairs of pure states. We construct explicit mutually inverse maps between observable-generated quadratic costs and Hilbert--Schmidt frame-type operators formed from the traceless parts of their generators. This correspondence also expresses the transport variational problem directly in terms of the frame operator. For every such cost $C$ on a $d$-dimensional Hilbert space, the minimal number of self-adjoint generators equals the rank of its recovered frame operator $\Theta(C)$, and is therefore at most $d^2-1$. Nonzero isotropic costs require exactly $d^2-1$ generators, attaining the largest minimal generator number that any observable-generated quadratic cost can require. Their arbitrary generating families give tight frames, while their minimal realizations give scaled Hilbert--Schmidt orthonormal bases of the traceless subspace. These symmetry, frame, and diametral characterizations determine the same one-parameter family of quantum Wasserstein distances.

\end{abstract}

\maketitle
\section{Introduction and main result}

Quantum optimal transport compares quantum states through transport-type variational principles and gives rise to questions about the geometry and symmetries of the state space. We refer to the book
\cite{maasOptimalTransportQuantum2024}
and to the survey papers
\cite{
beattyWassersteinDistancesQuantum2025,
trevisanQuantumOptimalTransportInvitation2025}
for overviews. A dynamical theory was developed by Carlen and Maas
\cite{
carlenMaasGradientFlow2014,
carlenMaasGradientFlowQuantum2017,
carlenMaasNoncommutativeCalculus2017};
see also the works of Datta and Rouzé
\cite{dattaRelatingRelativeEntropy2020}
and Wirth
\cite{
wirthDualFormulaNoncommutative2022}.
Caglioti, Golse, Mouhot, and Paul used quantum couplings to formulate quantum optimal transport problems and the corresponding Wasserstein distances
\cite{
cagliotiGolsePaulQuantumOptimalTransportCheaper2020,
cagliotiGolsePaulTowardsOptimal2023,
golsePaulWavePackets2018,
golseOptimalTransportPseudometrics2022,
golseMouhotPaulMeanFieldClassical2016},
whereas in the approach of De Palma and Trevisan
\cite{
depalmaQuantumOptimalTransport2021,
depalmaQuantumOptimalTransport2024}
the transport is realized by quantum channels; see also
\cite{
bunthMetricPropertyQuantum2024,
bunthStrongKantorovichDuality2026,
bunthSwapTransposeCouplings2025,
bunthWassersteinDistancesDivergences2025,
wirthTriangleInequalityQuantum2025}
for further developments in this direction. The construction of Friedland, Eckstein, Cole, and Życzkowski
\cite{
friedlandEcksteinColeZyczkowskiQuantumWasserstein,
coleEcksteinFriedlandZyczkowskiQuantumOptimalTransport2023}
is likewise coupling-based, but uses cost operators of a rather different nature. Duvenhage introduced quantum Wasserstein distances via modular couplings
\cite{
duvenhageOptimalQuantumChannels2021,
duvenhageQuadraticWassersteinMetrics2022},
and separable quantum Wasserstein distances have also been defined and investigated
\cite{
beattyOrderQuantumWasserstein2026,
tothPitrikQuantumWassersteinSeparable2023,
tothQuantumWassersteinDistance2026}.

The symmetries of these transport geometries are natural objects of study. Hermann Weyl wrote in his book
\emph{Symmetry}
\cite{weylSymmetry1989}:
\begin{quote}
Whenever you have to do with a structure--endowed entity $\Sigma$, try to determine its group of automorphisms, the group of those element--wise transformations which leave all structural relations undisturbed. You can expect to gain a deep insight into the constitution of $\Sigma$ in this way.
\end{quote}
For a Wasserstein geometry on a state space, Weyl's principle translates into the problem of identifying those transformations of states that preserve the induced Wasserstein distance. For genuine metrics, the bijective transformations of this kind form the isometry group of the metric space. The generalized Wasserstein distances considered below, however, need not be genuine metrics, and their distance-preserving transformations need not be bijective. The natural object is therefore the full isometry monoid.

While transformations preserving quantum-information divergences on state spaces and positive operator cones have been studied extensively
\cite{
molnarMapsPositiveDefinite2016,
virosztekMapsQuantumStates2016,
virosztekQuantumFdivergencePreserving2016},
and classical Wasserstein isometries have likewise attracted substantial attention
\cite{
geherIsometricStudyWasserstein2020,
geherIsometryGroupWasserstein2022},
the first works on quantum Wasserstein isometries appeared only very recently
\cite{
geherQuantumWassersteinIsometries2023,
simonIsometriesQubitState2025,
bunthQuantumWassersteinIsometries2026,
szaboQuantumWassersteinIsometries2026}.
The qubit results in \cite{geherQuantumWassersteinIsometries2023,simonIsometriesQubitState2025} concern several distinguished transport costs and their associated distances and divergences. For the cost generated by all three Pauli observables, the isometries are exactly the Wigner symmetries, that is, unitary and antiunitary conjugations. Removing one Pauli observable permits additional isometries, including non-surjective and non-injective ones \cite{geherQuantumWassersteinIsometries2023}. Isometries for costs generated by a single observable have also been studied in arbitrary finite dimension \cite{szaboQuantumWassersteinIsometries2026}. These distances need not be genuine metrics, so the usual metric argument for injectivity does not apply.

For the symmetric cost generated by tensor products of Pauli operators, the same Wigner characterization holds on every $n$-qubit state space \cite{bunthQuantumWassersteinIsometries2026}.
The present paper starts from this observation, but takes a dimension-independent and intrinsic point of view. Instead of fixing a distinguished cost or a particular number of generating observables, we ask which quadratic costs produce exactly the Wigner symmetries, and how this symmetry can be recognized from the geometry of the distance.

Let $\cH$ be a $d$-dimensional complex Hilbert space and let
$\cA=\crbr{A_1,\ldots,A_K}\subseteq\bsa{\cH}$ be any finite family of self-adjoint observables. Its quadratic transport cost operator is
\be\label{eq:intro-cost}
    C_{\cA}
    =
    \sum_{k=1}^K
    \rbr{A_k\otimes I^T-I\otimes A_k^T}^2,
\ee
and the corresponding quantum Wasserstein distance is defined by
\be\label{eq:intro-distance-squared}
    D^2_{\cA}\rbr{\rho,\omega}
    :=
    \inf_{\Pi\in\cC\rbr{\rho,\omega}}\Tr\Pi C_{\cA}.
\ee
Here $\cC\rbr{\rho,\omega}$ consists of the quantum states on
$\hoh$ with marginals $\omega$ and $\rho^T$. The notation and the underlying double-ket formalism are developed in \Cref{sec:background,sec:Wasserstein-intro}; until then we use them to describe the results.

Our analysis separates the geometry of the distance from the choice of its generators. First, for arbitrary positive semidefinite costs, the distance on pairs of pure states determines the cost operator completely. Unitary invariance of the pure-state distance therefore forces the cost to intertwine the adjoint representation. Observable-generated quadratic costs vanish on the scalar component. Since the adjoint representation is irreducible on the traceless component, Schur's lemma then forces every nonzero invariant cost of the form \eqref{eq:intro-cost} to be a positive scalar multiple of the identity on that component.

Second, we make the realization by observables intrinsic. To $\cA$ we associate the Hilbert--Schmidt frame-type operator
\be
    F_{\cA}
    :=
    \sum_{k=1}^K
    \kketbbra{\pi_0\rbr{A_k}}{\pi_0\rbr{A_k}},
    \qquad
    \pi_0(A):=A-\frac{\Tr A}{d}I.
\ee
We construct explicit forward and recovery maps $\Sigma$ and $\Theta$, which give mutually inverse bijections between these frame-type operators and observable-generated quadratic costs. Thus
$C_{\cA}=\Sigma(F_{\cA})$ and $F_{\cA}=\Theta(C_{\cA})$, independently of the chosen generating family. We also write the transport variational problem directly on the frame side: the marginal contribution is fixed by a partial trace of $F_{\cA}$, while the optimization concerns its pairing with the reshuffled coupling. For product couplings this reduces to a sum of two total variances and a quadratic separation measured by the frame operator.

This correspondence identifies the exact generator complexity of every observable-generated quadratic cost. Writing $m(C)$ for the smallest number of self-adjoint observables realizing $C$, we prove
\be
    m(C)=\rank\Theta(C)\leq d^2-1.
\ee
The upper bound is therefore universal: every such cost admits a realization with at most $d^2-1$ observables, even if the family initially used to define it is overcomplete. For a nonzero isotropic cost the recovered frame operator is a positive multiple of the identity on the traceless subspace, so
$m(C)=d^2-1$. In this precise sense, symmetric Wasserstein distances require the maximal number of generators that any observable-generated quadratic cost can require. Their generating families are exactly the tight Hilbert--Schmidt frames of the traceless parts with the prescribed frame constant; a minimal realization is a scaled Hilbert--Schmidt orthonormal basis. Maximal minimal generator number alone is not a characterization of symmetry: isotropy also requires equality of the nonzero frame eigenvalues.

Finally, \Cref{sec:main-theorem} gives a direct geometric characterization of isotropy of the cost operator in terms of diametral pairs. If
\be
    C=\beta P_{\kket{\cV(d)}},\qquad \beta>0,
\ee
then the Wasserstein diameter is $\sqrt{\beta}$, and two states $\rho,\omega$ attain it if and only if their supports are orthogonal, equivalently $\rho\omega=0$ or $\Tr\rho\omega=0$. In the orthogonal case, in fact, every coupling of $\rho$ and $\omega$ is optimal. Conversely, for a nonzero observable-generated cost, it is enough that every orthogonal pair of pure states attains the diameter, then the cost is isotropic. Thus the main theorem links Wigner symmetry, unitary invariance, isotropy, tight frames, minimal realizations, and diametral orthogonality.

The main result is stated below and restated and proved in \Cref{sec:main-theorem}.

\begin{maintheorempreview}
    Let $\cH$ be a $d$-dimensional Hilbert space with $d>1$, and let
    $\cA=\crbr{A_k}_{k=1}^K\subseteq\bsa{\cH}$ be a finite family of observables, with $K\geq1$ arbitrary. Let $C_{\cA}$ be its quadratic cost operator and $D_{\cA}$ its Wasserstein distance. The following statements are equivalent.
    \bi
        \ii The $D_{\cA}$-isometry monoid is precisely the set of Wigner symmetries, that is, the conjugations by unitary or antiunitary operators.

        \ii $D_{\cA}$ is not identically zero and its restriction to pure states is invariant under unitary conjugations.

        \ii $C_{\cA}\neq0$ and the cost is invariant under the adjoint unitary representation:
        \be
            \Ad_U^*C_{\cA}\Ad_U
            =
            \rbr{U\otimes\bar U}^*C_{\cA}\rbr{U\otimes\bar U}
            =
            C_{\cA},
            \qquad U\in\bu{\cH}.
        \ee

        \ii The cost is fully isotropic on the traceless component: there exists $\alpha>0$ such that
        \be
            C_{\cA}
            =
            2d\alpha
            \rbr{
                I\otimes I^T-\frac1d\kketbbra{I}{I}
            }.
        \ee

        \ii The traceless parts of the given generators form a Hilbert--Schmidt tight frame: there exists $\alpha>0$ such that
        \be
            F_{\cA}
            =
            \sum_{k=1}^K\kketbbra{\pi_0\rbr{A_k}}{\pi_0\rbr{A_k}}
            =
            \alpha
            \rbr{
                I\otimes I^T-\frac1d\kketbbra{I}{I}
            }.
        \ee

        \ii $m\rbr{C_{\cA}}=d^2-1$ and a family $\cB=\crbr{B_j}_{j=1}^{d^2-1}\subseteq\bsa{\cH}$ realizes the cost, $C_{\cA}=C_{\cB}$, if and only if there exists $\alpha>0$ such that
        \be
            \crbr{
                \frac{1}{\sqrt{\alpha}}\pi_0\rbr{B_j}
            }_{j=1}^{d^2-1}
        \ee
        is a Hilbert--Schmidt orthonormal basis of the complex traceless subspace of $\bo{\cH}$.

        \ii There exists $\alpha>0$ such that
        \be
            \sup_{\rho,\omega\in\cS\rbr{\cH}}D_{\cA}\rbr{\rho,\omega}
            =\sqrt{2d\alpha},
        \ee
        and for all states $\rho,\omega\in\cS\rbr{\cH}$,
        \be
            D_{\cA}\rbr{\rho,\omega}
            =
            \sup_{\rho',\omega'\in\cS\rbr{\cH}}D_{\cA}\rbr{\rho',\omega'}
            \quad\iff\quad
            \rho\omega=0.
        \ee

        \ii There exists $\alpha>0$ such that
        \be
            \sup_{\rho,\omega\in\cS\rbr{\cH}}D_{\cA}\rbr{\rho,\omega}
            =\sqrt{2d\alpha},
        \ee
        and for all pure states $P,Q\in P_1\rbr{\cH}$,
        \be
            PQ=0\quad\Longrightarrow\quad D_{\cA}\rbr{P,Q}
            =
            \sup_{\rho,\omega\in\cS\rbr{\cH}}D_{\cA}\rbr{\rho,\omega}.
        \ee
    \ei
    Whenever these conditions hold all $\alpha$ parameters above coincide.
    Furthermore, $D_{\cA}$ is not identically zero if and only if $C_{\cA}\neq0$, which in turn holds if and only if at least one $\pi_0(A_k)$ is nonzero.
\end{maintheorempreview}

\br
    The aim of \Cref{thm:main} is two-fold.
    \bi
        \ii It characterizes symmetric Wasserstein distances on arbitrary finite-dimensional Hilbert spaces, extending the previously known cases
        \cite{bunthQuantumWassersteinIsometries2026}.

        \ii It removes the ambiguity in the term ``symmetric Wasserstein distance'' and makes it intrinsic. Indeed, the isometric, representation-theoretic, operator-theoretic, frame-theoretic and diametric symmetry properties listed in \Cref{thm:main} are equivalent and determine the same Wasserstein distance, up to a multiplicative constant in all finite dimensions.
    \ei
\er

The paper is organized as follows. \Cref{sec:background} develops the basis-independent Dirac and double-ket formalism and recalls the required representation theory. \Cref{sec:Wasserstein-intro} reviews quantum couplings, studies one- and two-sided quadratic costs, and proves that a positive semidefinite cost is determined by its pure-state distance. \Cref{sec:frame-characterization} establishes the frame--cost correspondence, uses it to evaluate transport costs, and identifies the minimal generator number and the isotropic costs. \Cref{sec:main-theorem} establishes both directions of the diametral characterization and proves the main theorem. \Cref{app:recast-proof} records the dimension-independent version of the Wigner-isometry argument from an earlier joint work.

\section{Notations and mathematical background}\label{sec:background}
$\sqbr{K}:=\crbr{1,\ldots,K}$, with $[0]=\varnothing$ if $K=0$. Throughout this paper all Hilbert spaces are complex and finite-dimensional, unless stated otherwise. $\Lin\rbr{\cH,\cK}$ denotes the linear operators from the Hilbert space $\cH$ to the Hilbert space $\cK$. We use the shorthand $\bo{\cH}=\Lin\rbr{\cH,\cH}$. $\bsa{\cH}$, $\bu{\cH}$, $P_1\rbr{\cH}$ denote the selfadjoint operators, the unitary group and the set of rank-one projections, or equivalently pure states over the Hilbert space $\cH$ respectively. $P_V$ denotes the orthogonal projection onto the subspace $V$, where the ambient Hilbert space should be clear from context. $\cS\rbr{\cH}$ denotes the set of quantum states over $\cH$, i.e. the set of positive, trace-one operators. $\pi_0$ denotes the projection onto the traceless part, that is, the difference of the identity map and the completely depolarizing channel,
\be
    \pi_{0}\rbr{A}=A-\Tr A\frac{I}{d},\qquad A\in\bo{\cH}.
\ee
We first define kets and double kets as linear maps and specify the canonical identifications used below.
\subsection{Dirac formalism}
\bde
    Let $\cH$ be a finite-dimensional complex Hilbert space, with inner product complex linear in its second variable. The canonical linear ket map and conjugate-linear bra map are
    \be
        \ket{\cdot}:\cH\to\Lin(\bC,\cH),
        &\qquad \ket{\psi}(\lambda):=\lambda\psi,\\
        \bra{\cdot}:\cH\to\Lin(\cH,\bC),
        &\qquad \bra{\psi}(\varphi):=(\psi,\varphi).
    \ee
\ede
\br
    In this paper, we use the Dirac formalism. Under the Dirac formalism the vectors $\psi\in\cH$ are identified with the linear maps $\ket{\psi}$, linear functionals are given by the linear maps $\bra{\psi}$, complex numbers are identified with elements of $\Lin(\bC,\bC)$ and for any linear operator $A\in \Lin\rbr{\cH_1,\cH_2}$ between finite-dimensional Hilbert spaces $\cH_1,\cH_2$
    \be
        A\in\span\rbr{\ketbra{\psi}{\varphi}:\psi\in \cH_2, \varphi\in\cH_1}.
    \ee
    With the Dirac notation in mind, vectors, covectors, and even complex numbers are regarded as linear maps. Accordingly, all inner products appearing in this formalism may be viewed as Hilbert--Schmidt inner products on appropriate spaces of linear maps, and the result of the trace is regarded as an element of $\Lin(\bC,\bC)$, rather than a bare complex number. In this convention, when a Hilbert-space operator is identified with the corresponding operator between ket spaces, its adjoint is the Hilbert--Schmidt adjoint of that induced map. This recovers the usual Hilbert-space adjoint under the identification
    $\cH\simeq\Lin(\bC,\cH)$. Indeed, it is easy to see that $\ket{\psi}^*=\bra{\psi}$, for every $\psi\in\cH$, and hence
    \be
            \Tr \ket{\psi}^*\ket{\varphi}=\Tr \bra{\psi}{}\ket{\varphi}=\bra{\psi}{}\ket{\varphi}=:\braket{\psi}{\varphi}:\lambda\mapsto\lambda \rbr{\psi,\varphi},\quad \forall\psi,\varphi\in\cH,\;\forall \lambda\in\bC.
    \ee
\er

We will make use of the complex polarization identity.
\bl[Complex polarization]\label{lem:polarization}
    Let $V$ and $W$ be complex vector spaces, and let
    $\gamma:V\times V\to W$ be sesquilinear, that is conjugate-linear in the first variable and linear in the second variable. Then
    \be
        \gamma(x,y)
        =
        \frac{1}{4}\sum_{k=1}^{4} i^k\,
        \gamma\!\rbr{i^k x+y,i^k x+y},
        \qquad
        x,y\in V.
    \ee
\el

We will also make use of the following lemma characterizing frame-operators.

\bl[Frame operators]\label{lemma:sum-of-rank-ones}
    Let $V$ be a $d$-dimensional Hilbert space over $\bC$, and let
    $v_1,\dots,v_K\in V$. Suppose that there exists
    $\alpha>0$ such that
    \be\label{eq:frame-oeprator}
        \sum_{k=1}^K \ketbra{v_k}{v_k}=\alpha I_{V}.
    \ee
    Then
    \be
        K\geq d.
    \ee
    Moreover, if $K=d$, then \eqref{eq:frame-oeprator} is satisfied if and only if $\crbr{\frac{1}{\sqrt{\alpha}} v_k}_{k=1}^d$ forms an ONB of $V$.
\el

\bpr
    The rank of $\sum_k\ketbra{v_k}{v_k}$ is at most $K$, so $K\geq d$.
    If $K=d$, let $A:\bC^d\to V$ be the linear map $Ae_k=v_k$.
    Then \eqref{eq:frame-oeprator} says $AA^*=\alpha I_V$.
    Since $A$ is square, this is equivalent to $A^*A=\alpha I_{\bC^d}$,
    or $\braket{v_j}{v_k}=\alpha\delta_{jk}$, as required.
\epr

\bde
    For a linear map $A:\cH_1\to \cH_2$, we denote by
    $A^T:\cH_2^*\to \cH_1^*$ its algebraic transpose, defined by
    \be
        A^T(\eta)=\eta\circ A,\qquad \eta\in \cH_2^*.
    \ee
    We will use the shorthand
    \be
        \bar A := (A^*)^T.
    \ee
\ede

\bde\label{def:doubleket}
    Let $\cH$ be a finite-dimensional Hilbert space. We define the canonical linear map
    \be
        \kket{\cdot} : \bo{\cH} \to \Lin  (\bC,\cH)\otimes\Lin (\bC^*,\cH^*)
    \ee
    as the linear extension of
    \be
        \ket{\psi}\bra{\varphi} \mapsto \ket{\psi} \otimes \bra{\varphi}^T,
        \qquad \rbr{\psi,\varphi \in \cH}.
    \ee
\ede

\bp\label{prop:doubleket-expansions}
    Let $\cH$ be a finite-dimensional Hilbert space, let
    $A\in\bo{\cH}$, and let $\cE=\crbr{e_i}$ and
    $\cF=\crbr{f_j}$ be orthonormal bases of $\cH$. Let
    \be
        A=\sum_{\ell=1}^M c_\ell\ketbra{a_\ell}{a_\ell},
        \qquad
        c_\ell\in\bC,\quad a_\ell\in\cH,
    \ee
    be any decomposition of $A$ into self-dyads with complex coefficients, as supplied by complex polarization. Then
    \be\label{eq:doubleket-expansion}
        \kket{A}
        &=\sum_{\ell=1}^M c_\ell\ket{a_\ell}\otimes\bra{a_\ell}^T
        =\sum_{i,j}^{\dim \cH}\braket{e_i}{Af_j}\ket{e_i}\otimes\bra{f_j}^T\\
        &=\sum_j^{\dim \cH}\ket{Af_j}\otimes\bra{f_j}^T
        =\sum_i^{\dim \cH}\ket{e_i}\otimes\bra{A^*e_i}^T.
    \ee
    Furthermore,
    \be\label{eq:doublebra-expansion}
        \bbra{A}:=\kket{A}^*
        &=\sum_{\ell=1}^M\overline{c_\ell}\bra{a_\ell}\otimes\ket{a_\ell}^T
        =\sum_{i,j}^{\dim \cH}\braket{f_j}{A^*e_i}\bra{e_i}\otimes\ket{f_j}^T\\
        &=\sum_j^{\dim \cH}\bra{Af_j}\otimes\ket{f_j}^T
        =\sum_i^{\dim \cH}\bra{e_i}\otimes\ket{A^*e_i}^T.
    \ee
    In particular, the map $\kket{\cdot}$ is well defined and independent both of the self-dyad representation and of the orthonormal bases used in these expansions.
\ep

\bpr
    The first two equalities in \eqref{eq:doubleket-expansion} follow from linearity, the last two equalities by the definition of the adjoint and dropping a resolution of the identity. \eqref{eq:doublebra-expansion} is then just the adjoint of \eqref{eq:doubleket-expansion}.
\epr

\bc \label{cor:computational-rules}
    For all $A,B,C\in\bo{\cH}$, the following identities hold:
    \bi
        \ii \be \label{eq:product-rule}
            (B \otimes C^T)\kket{A} = \kket{BAC},
        \ee
        \ii \be \label{eq:product-rule2}
            \bbra{A}(B^* \otimes \bar C) = \bbra{BAC},
        \ee
        \ii \be \label{eq:kket-isomorphism}
            \Tr \kket{A}^*\kket{B}=\bbrakket{A}{B} = \bbra{I}(A^*\otimes B^T)\kket{I} = \Tr A^* B,
        \ee
        \ii \be \label{eq:partial-traces-identity}
            \Tr_{\cH^*} \kketbbra{A}{B}=AB^*,
            \qquad
            \Tr_{\cH} \kketbbra{A}{B}=\rbr{B^*A}^T=A^T\bar B.
        \ee
        \ii For every state $\rho=\sum_i\lambda_i\ketbra{r_i}{r_i}$, with $\crbr{r_i}$ an orthonormal eigenbasis,
        \be \label{eq:canonical-purification}
            \kketbbra{\sqrt{\rho}}{\sqrt{\rho}}
            = \sum_{i,j}\sqrt{\lambda_i\lambda_j}
            \ket{r_i}\bra{r_j}\otimes\rbr{\ket{r_j}\bra{r_i}}^T.
        \ee
        This is a basis-independent purified state with marginals $\rho$ and $\rho^T$, called the canonical purification of $\rho$.
    \ei
\ec

\bpr
    The identities follow from \Cref{prop:doubleket-expansions}.
\epr

\br
    We emphasize that Dirac notation is used in the paper to keep track of the underlying mapping structure. Thus $\ket{\psi}$ is first viewed as an element of $\Lin(\bC,\cH)$, rather than identified a priori with a vector in $\cH$. In the sequel, however, we will also freely use the standard identifications
    \be
        \cH\simeq \Lin(\bC,\cH),\qquad
        \cH^*\simeq \Lin(\bC^*,\cH^*),\qquad
        \hoh
        \simeq
        \Lin(\bC,\cH)\otimes\Lin(\bC^*,\cH^*),
    \ee
    in addition to the identification
    \be
    \bo{\cH}\simeq \Lin(\bC,\cH)\otimes\Lin(\bC^*,\cH^*), 
    \ee
    introduced by \Cref{def:doubleket}.
    Under these identifications, $\kket{A}$ is the ket vector, $\ket{A}$, corresponding to $A\in\bo{\cH}$, but with the additional advantage that left and right multiplication are encoded by \eqref{eq:product-rule} and \eqref{eq:product-rule2}.
\er

By complex polarization, rank-one projections span $\bo{\cH}$. Their tensor products therefore give the following separation properties.

\bp\label{prop:doubleket-categories}\phantom{1}
    \bi
        \ii\label{item:doublket-categories1} For any $C\in \bo{\hoh}$
        \be\label{eq:general-hoh-decomp}
            C=\sum_{i} c_{i}\kket{\ketbra{x_i}{y_i}}\bbra{\ketbra{x_i}{y_i}}=\sum_i c_i\ketbra{x_i}{x_i}\otimes \rbr{\ketbra{y_i}{y_i}}^T,
        \ee
        for appropriately chosen complex coefficients $c_{i}$ and $x_i,y_i\in \cH$. The vectors may be chosen of unit length after absorbing norms into the coefficients.
        \ii\label{item:doublket-categories2} For any $C\in \bo{\hoh}$, $C=0$ if and only if
        \be
            \bbra{\ketbra{x}{y}}C\kket{\ketbra{x}{y}}=\rbr{\bra{x}\otimes \ket{y}^T}C\rbr{\ket{x}\otimes \bra{y}^T}=0,
            \qquad
            x,y\in\cH.
        \ee
        \ii\label{item:doublket-categories3} $C$ is selfadjoint if and only if
        the coefficients $c_{i}$ in \eqref{eq:general-hoh-decomp} can be chosen to be reals.
    \ei
\ep
\bpr
    Rank-one projections span $\bo{\cH}$ by complex polarization, so their
    tensor products span $\bo{\hoh}$. This proves
    \ref{item:doublket-categories1}.
    If the quadratic forms in \ref{item:doublket-categories2} vanish,
    then $\Tr(CX)=0$ on this spanning family and hence for every
    $X\in\bo{\hoh}$; thus $C=0$.
    Finally, if $C=C^*$, averaging \eqref{eq:general-hoh-decomp} with its
    adjoint replaces each coefficient $c_i$ by $\Re c_i$.
    The converse in \ref{item:doublket-categories3} is immediate.
\epr

\subsection{Representation theoretic background}
In this subsection we recall the minimal representation-theoretic notation and facts needed for the treatment of Wasserstein quadratic cost operators.
\bde
    Let $S \in \bo{\cH}$ be invertible and define
    \be
        \Ad_S : \bo{\cH} \to \bo{\cH}, \qquad\Ad_S(B) := S B S^{-1}.
    \ee
    This defines the adjoint action of the invertible group on $\bo{\cH}$. In the double ket formalism the same action is identified as
    \be
        \Ad_S : \hoh \to \hoh, \qquad
        \Ad_S\kket{B}:= \kket{\Ad_S (B)}=S\otimes \rbr{S^{-1}}^T\kket{B},
    \ee
    with the last equality following from \eqref{eq:product-rule}.
    Restricting to $\bu{\cH}$ gives the adjoint representation of $\bu{\cH}$ on $\hoh$ and equivalently on $\bo{\cH}$:
    \be
        \Ad_U(B)=UBU^*,\qquad\;\Ad_U \kket{B}=U\otimes \bar{U} \kket{B}.
    \ee
\ede

\bde
    Let $X \in \bo{\cH}$. The linear map
    \be
        \ad_X : \bo{\cH} \to \bo{\cH}, \qquad
        \ad_X(B) := \sqbr{X,B}= XB - BX
    \ee
    is called the infinitesimal adjoint action generated by $X$. In the double ket formalism the same action is identified as
    \be
        \ad_X : \hoh \to \hoh, \qquad
        \ad_X\kket{B}:= \kket{\ad_X (B)}=\rbr{X\otimes I^T-I\otimes X^T}\kket{B},
    \ee
    with the last equality following from \eqref{eq:product-rule}.
    With this convention no factor of $i$ is inserted; thus for self-adjoint $X$, the map $\ad_X$ is self-adjoint with respect to the Hilbert--Schmidt inner product.
\ede

\bp\label{prop:Ad-basic-identities}
    Let $S\in\bo{\cH}$ be invertible then
    \be
        \Ad_S^{-1}=\Ad_{S^{-1}},\quad        \Ad_S^*=\Ad_{S^*}.
    \ee
    In particular, if $U\in\bu{\cH}$ is unitary, then
    \be
        \Ad_U^*=\Ad_U^{-1}=\Ad_{U^*}.
    \ee
\ep

\bp\label{prop:ad-basic-identities}
    Let $X,Y\in\bo{\cH}$, then
    \be
        \ad_X^*=\ad_{X^*},\quad      \sqbr{\ad_X,\ad_Y}=\ad_{\sqbr{X,Y}}.
    \ee
\ep
\bde
    For $\cH\simeq\bC^d$ let us define the complex Hilbert spaces
    \be
        \cV_0(d)&:=\crbr{\lambda I_\cH\,:\,\lambda\in\bC}\simeq\crbr{\lambda\kket{I_\cH}\,:\,\lambda\in\bC}=:\kket{\cV_0(d)},\\  
        \cV(d)&:=\crbr{B\in\bo{\cH}\,:\,\Tr(B)=0}\simeq\crbr{\kket{B}\,:\,B\in\bo{\cH},\,\Tr(B)=0}=:\kket{\cV(d)}.
    \ee
    $\cV_0(d)$ and $\kket{\cV_0(d)}$ are $1$ dimensional, $\cV(d)$ and $\kket{\cV(d)}$ are $d^2-1$ dimensional complex Hilbert spaces. Let us also define the real Hilbert spaces
    \be
        \widetilde\cV(d):=\crbr{B\in\bo{\cH}\,:\,B=B^*,\,\Tr(B)=0}\simeq\crbr{\kket{B}\,:\,B\in\bo{\cH},\,B=B^*,\,\Tr(B)=0}=:\kket{\widetilde\cV(d)},
    \ee
    of real dimension $d^2-1$.
\ede

\bp\label{prop:irreducibility}
    Let $\cH\simeq\bC^d$, $d\geq2$. Under unitary conjugation,
    \be
        \bo{\cH}
        =
        \bC I
        \oplus
        \crbr{A\in\bo{\cH}:\Tr A=0},
    \ee
    and both summands are irreducible. Equivalently,
    \be
        \bo{\cH}
        =
        \cV_0(d)
        \oplus
        \cV(d)
    \ee
    is the decomposition of the representation
    $U\mapsto\Ad_U$ into irreducible components.

    For $d=1$, $\cV(1)=\crbr{0}$ and
    $\bo{\bC}=\cV_0(1)$ is irreducible.
\ep

\bpr
    Unitary conjugation fixes the identity and preserves the trace, so both
    summands are invariant. The one-dimensional subspace
    $\bC I=\cV_0(d)$ is trivially irreducible.

    It remains to consider the traceless subspace
    \be
        \cV(d)\simeq\mathfrak{sl}_d(\bC).
    \ee
    Scalar unitaries act trivially by conjugation, so it suffices to restrict
    the representation to $\SU(d)$. Under the above identification,
    this is the complexified adjoint representation of $\SU(d)$.
    Since $\mathfrak{sl}_d(\bC)$ is a simple Lie algebra, its adjoint
    representation is irreducible: its invariant subspaces are precisely its
    ideals. The equivalence between irreducibility for a connected Lie group,
    its Lie algebra, and the complexified Lie algebra, together with the
    simplicity of $\mathfrak{sl}_d(\bC)$, may be found in
    \cite[Props.~4.5--4.6 and Thm.~8.51]{hallLieGroupsLie2015}.
    Hence $\cV(d)$ is irreducible.
\epr

\bl[Schur's lemma]\label{lem:Schur}
    Let $(\pi,V)$ and $(\sigma,W)$ be finite-dimensional irreducible representations of a group $G$ over an algebraically closed field $\bF$. Let
    \be
        T:V\to W
    \ee
    be an intertwiner, i.e.
    \be
        T\pi(g)=\sigma(g)T
        \qquad \rbr{g\in G}.
    \ee
    Then:
    \bi
        \ii either $T=0$ or $T$ is an isomorphism;
        \ii if $V$ and $W$ are not isomorphic as representations, then $T=0$;
        \ii if $V=W$ and $\pi=\sigma$, then
        \be
            T=\lambda I
        \ee
        for some $\lambda\in\bF$.
    \ei
\el

\bc\label{cor:decomposition}
    Every linear map $T\in\bo{\hoh}$ satisfying
    \be
        T\Ad_U=\Ad_UT
        \qquad \rbr{U\in \bu{\cH}}
    \ee
    or equivalently
    \be
        \Ad_U^*T\Ad_U=T
        \qquad \rbr{U\in \bu{\cH}}
    \ee
    acts as a scalar on each irreducible component, $\kket{\cV(d)}$ and $\kket{\cV_0(d)}$. Hence there exist $\alpha,\beta\in\bC$ such that
    \be
        T
        =
        \alpha P_{\kket{\cV_0(d)}}
        +
        \beta P_{\kket{\cV(d)}}=\alpha \frac1d\kketbbra{I}{I}+\beta \rbr{I\otimes I^T-\frac1d\kketbbra{I}{I}}.
    \ee
\ec
\section{Wasserstein quadratic costs and distances}\label{sec:Wasserstein-intro}

We recall the channel-based formulation of quantum optimal transport introduced by De Palma and Trevisan in \cite{depalmaQuantumOptimalTransport2021}. In the classical theory, a transport plan between two probability measures may be viewed equivalently as a coupling, or as a stochastic map which sends the source distribution to the target distribution. The main idea of \cite{depalmaQuantumOptimalTransport2021} is to impose the same operational interpretation in the quantum setting: transport plans should be quantum channels.

Let $ \rho,\omega \in \cS\rbr{\cH} $. In \cite{depalmaQuantumOptimalTransport2021} a quantum transport plan from $ \rho $ to $ \omega $ is defined as a quantum channel
\be
    \Phi:
   \bo{\ran\rbr{\rho}}
    \to
    \bo{\cH}
\ee
such that
\be
    \Phi\rbr{\rho} = \omega.
\ee
Such a channel determines a state on $\hoh$ by applying $ \Phi $ to the first subsystem of the canonical purification of $ \rho $ (see \eqref{eq:canonical-purification}):
\be \label{eq:Pi-phi-def}
    \Pi_{\Phi}
    =
    \rbr{
    \Phi \otimes \id_{\bo{\cH^*}}
    }
    \rbr{
    \kket{\sqrt{\rho}}\bbra{\sqrt{\rho}}
    }.
\ee
The basic identities for the canonical vectorization in \Cref{cor:computational-rules} imply
\be
    \Tr_{\cH^*}\sqbr{\Pi_{\Phi}}
    =
    \omega,
    \qquad
    \Tr_{\cH}\sqbr{\Pi_{\Phi}}
    =
    \rho^T.
\ee
This motivates the following definition of quantum couplings.
\bde
The set of couplings with respect to the ordered pair $\rho,\omega$ is defined as
    \be \label{eq:q-coup-def}
        \cC\rbr{\rho,\omega}
        =
        \crbr{
        \Pi \in \cS\rbr{\hoh}
        \middle|
        \Tr_{\cH^*}\sqbr{\Pi} = \omega,
        \Tr_{\cH}\sqbr{\Pi} = \rho^T
        }.
    \ee
\ede
\br\label{rem:pure-couplings}
    If either $\rho$ or $\omega$ is pure, then
    \be
        \cC(\rho,\omega)=\crbr{\omega\otimes\rho^T}.
    \ee
    This follows from the elementary fact that a bipartite state with one pure
    marginal is necessarily the tensor product of its marginals.
\er
The assignment $ \Phi \mapsto \Pi_{\Phi} $ gives a one-to-one correspondence between quantum transport plans from $ \rho $ to $ \omega $ and quantum couplings in $ \cC\rbr{\rho,\omega} $. Thus the coupling formulation and the channel formulation are equivalent, while the latter keeps the interpretation of transport as a physical operation.

Let now
\be
    \cA = \crbr{A_1,\dots,A_K}
\ee
be a finite family of observables on $ \cH $. The corresponding quadratic cost operator in \cite{depalmaQuantumOptimalTransport2021} is defined as
\be \label{eq:quadratic-cost-operator}
    C_{\cA}
    =
    \sum_{k=1}^K
    \rbr{
    A_k \otimes I^T
    -
    I \otimes A_k^T
    }^2
    \in
    \bo{\hoh}.
\ee
The cost of a coupling $ \Pi \in \cC\rbr{\rho,\omega} $ is defined in \cite{depalmaQuantumOptimalTransport2021} by
\be \label{eq:cost-in-fin-dim-and-quad-cost-op}
    \Cost_{\cA}\rbr{\Pi}
    =
    \Tr_{\hoh}\sqbr{\Pi C_{\cA}}.
\ee
Equivalently, if $ \Pi = \Pi_{\Phi} $ is the coupling associated with a channel $ \Phi $, then the same cost can be written directly in terms of the channel as
\be \label{eq:cost-in-terms-of-channel}
    \Cost_{\cA}\rbr{\Pi_{\Phi}}
    =
    \sum_{k=1}^K
    \rbr{
    \Tr_{\cH}\sqbr{\Phi\rbr{\rho} A_k^2}
    +
    \Tr_{\cH}\sqbr{\rho A_k^2}
    -
    2\Tr_{\cH}\sqbr{
    \sqrt{\rho} A_k \sqrt{\rho}\Phi^*\rbr{A_k}
    }
    }.
\ee
Here
\be
    \Phi^*:
    \bo{\cH}
    \to
    \bo{\ran\rbr{\rho}}
\ee
denotes the adjoint of $ \Phi $, characterized by
\be
    \Tr_{\cH}\sqbr{\Phi\rbr{X}A}
    =
    \Tr_{\ran\rbr{\rho}}\sqbr{X\Phi^*\rbr{A}}
\ee
for all $ X \in \bo{\ran\rbr{\rho}} $ and $ A \in \bo{\cH} $.

Finally the quadratic quantum Wasserstein distance associated with the family of observables $ \cA = \crbr{A_1,\dots,A_K} $ is defined in \cite{depalmaQuantumOptimalTransport2021} as the function $ D_{\cA} $ on $ \cS\rbr{\cH} \times \cS\rbr{\cH} $ defined by
\be \label{eq:quadratic-QW-dist-def}
    D_{\cA}^2\rbr{\rho,\omega}
    =
    \inf
    \crbr{
    \Cost_{\cA}\rbr{\Pi}
    \middle|
    \Pi \in \cC\rbr{\rho,\omega}
    }.
\ee
Equivalently, the infimum may be taken over all quantum channels $ \Phi:\bo{\ran\rbr{\rho}} \to \bo{\cH} $ satisfying $ \Phi\rbr{\rho}=\omega $, with $ \Pi $ replaced by $ \Pi_{\Phi} $.

In the present paper we restrict ourselves to finite dimensions and there we take a slightly more general point of view than that of \cite{depalmaQuantumOptimalTransport2021}. Our construction is still based on their definition of quantum couplings, but we allow slightly generalized quadratic cost operators. The presentation is organized more explicitly around the computational rules of \Cref{cor:computational-rules} and the representation-theoretic interpretation that these rules make available, rather than around the identification of couplings with quantum channels.

\bde\label{def:one-sided-quadratic-cost}
    Let $\cH$ be a finite-dimensional Hilbert space and let $\cX:=\crbr{X_k}_{k=1}^K$ be a finite collection of arbitrary operators on $\cH$. The associated one-sided quadratic cost operator is the linear map
    \be\label{eq:one-sided-quadratic-cost-operator}
        C_\cX : \hoh \to \hoh,
        \qquad
        C_\cX = \sum_k \ad_{X_k}^*\ad_{X_k}=\sum_k (X_k^* \otimes I^T - I \otimes \bar{X_k})(X_k \otimes I^T - I \otimes X_k^T).
    \ee
\ede

\bde\label{def:two-sided-quadratic-cost}
    Let $\cH$ be a finite-dimensional Hilbert space and let $\cX:=\crbr{X_k}_{k=1}^K$ be a finite collection of arbitrary operators on $\cH$. The associated two-sided quadratic cost operator is the linear map
    \be\label{eq:two-sided-quadratic-cost-operator}
        \widetilde{C}_\cX &: \hoh \to \hoh,
        \qquad
        \widetilde{C}_\cX = \sum_k \frac{1}{2}\rbr{\ad_{X_k}^*\ad_{X_k}+\ad_{X_k}\ad_{X_k}^*}\\
        &=\frac{1}{2}\sum_k\rbr{ (X_k^* \otimes I^T - I \otimes \bar{X_k})(X_k \otimes I^T - I \otimes X_k^T)+(X_k \otimes I^T - I \otimes X_k^T)(X_k^* \otimes I^T - I \otimes \bar{X_k})}.
    \ee
    The adjective two-sided refers to the averaging of the two natural products  $\ad_{X_k}^*\ad_{X_k}$ and $\ad_{X_k}\ad_{X_k}^*$.
\ede

\br
    If every $X_k$ is selfadjoint, then $\ad_{X_k}^*=\ad_{X_k}$, so both
    definitions reduce to \eqref{eq:quadratic-cost-operator}.
    \be
        \bbra{B} C_{\cX}\kket{B}=\sum_k\|\sqbr{X_k,B}\|_{HS}^2,
    \ee
    proves positivity of the one-sided cost operator.

    Write $X=X_S+iX_A$, where $X_S=(X+X^*)/2$ and
    $X_A=(X-X^*)/(2i)$ are selfadjoint. By \Cref{prop:ad-basic-identities},
    \be\label{eq:nonsymmetric-terms}
        \ad_X^*\ad_X
        =\ad_{X_S}^2+\ad_{X_A}^2+\ad_{i\sqbr{X_S,X_A}}.
    \ee
    The first two terms preserve selfadjointness, whereas the last maps
    selfadjoint operators to skew-adjoint operators. Since $i\sqbr{X_S,X_A}$
    is traceless, its commutator map vanishes exactly when $\sqbr{X_S,X_A}=0$.
    Thus the one-sided cost operator of a single nonnormal generator is not observable-generated.
    For several generators these additional commutator terms may cancel.

    On the other hand,
    \be\label{eq:ymmetric-terms}
        \frac12\rbr{\ad_X^*\ad_X+\ad_X\ad_X^*}
        =\ad_{X_S}^2+\ad_{X_A}^2.
    \ee
    Hence two-sided cost operators are exactly the observable-generated cost operators:
    one arbitrary generator combines two selfadjoint generators.
    For $\cX^*:=\crbr{X_k^*}_{k=1}^K$, the two-sided cost operator is invariant under
    $\cX\mapsto\cX^*$, unlike the one-sided cost operator in general. Below we relate
    this invariance to symmetry of the distance in its two variables.
\er
\br \label{rem:selfadjoint-quadratic-cost}
    If $C$ is a quadratic cost operator generated by a collection of selfadjoint observables $\cA:=\crbr{A_k}_{k=1}^K$, then the one- and two-sided definitions coincide and we can write without ambiguity $C:=C_\cA$ as in \eqref{eq:quadratic-cost-operator}.
\er
\bc\label{cor:cost-maps-to-traceless}
    For a finite-dimensional Hilbert space $\cH$, the one-sided and two-sided quadratic cost operators \eqref{eq:one-sided-quadratic-cost-operator} and \eqref{eq:two-sided-quadratic-cost-operator} vanish on $\cV_0(d)$ and $\cV(d)$ is invariant under them.
\ec
\bpr
    Both cost operators annihilate $\kket{I}$ because
    $\ad_{X_k}(I)=\ad_{X_k^*}(I)=0$.
    Being selfadjoint, they preserve its orthogonal complement
    $\kket{\cV(d)}$.
\epr

\bp\label{prop:generalized-cost-gauge-invariance}
Let $\cH$ be a finite-dimensional Hilbert space and let $\crbr{X_k}_{k=1}^K \subseteq \bo{\cH}$ be a collection of operators. For each $k$, let $\alpha_k \in \bC$ and define
\be
    \widetilde X_k := X_k + \alpha_k I.
\ee
then
\be
    \ad_{\widetilde X_k}=\ad_{X_k}
\ee
and consequently the quadratic cost operators defined in \Cref{def:one-sided-quadratic-cost} and \Cref{def:two-sided-quadratic-cost} are invariant under the map $\crbr{X_k}_{k=1}^K\mapsto \crbr{\widetilde X_k}_{k=1}^K$.
\ep

\bpr
    The identity $\sqbr{X_k+\alpha_k I,B}=\sqbr{X_k,B}$ gives
    $\ad_{\widetilde X_k}=\ad_{X_k}$; both cost operators therefore remain unchanged.
\epr

\bde
    We define the cost of arbitrary $\Pi\in\cS\rbr{\hoh}$ with respect to arbitrary positive semidefinite $C\in\bo{\hoh}$ as
    \be \label{eq:cost-general}
        \Cost_C\rbr{\Pi}
        =
        \Tr_{\hoh}\sqbr{\Pi C}.
    \ee
    If $C$ is a quadratic cost operator generated by a collection of selfadjoint observables $\cA:=\crbr{A_k}_{k=1}^K$, then the one- and two-sided definitions coincide and we can write without ambiguity $\Cost_C:=\Cost_\cA$ as in \eqref{eq:cost-in-fin-dim-and-quad-cost-op}.
\ede

\bde \label{def:QW-distances}
    The quantum Wasserstein distance of the states $\rho,\omega \in \cS\rbr{\cH}$ with respect to any positive semidefinite cost operator $C$ on $\hoh$ is denoted by $D_{C}$ and is defined as the square root of the minimal cost on the couplings of $\rho$ and $\omega$ with respect to the cost operator $C$:
    \be\label{eq:QW-distances}
        D_{C}^2\rbr{\rho,\omega}
        =
        \inf
        \crbr{
            \Cost_C\rbr{\Pi}
            \,\middle|\,
            \Pi \in \cC\rbr{\rho,\omega}
        }.
    \ee
    If the cost operator $C$ is a quadratic cost operator generated by a collection of selfadjoint observables $\cA:=\crbr{A_k}_{k=1}^K$, then we can write without ambiguity $D_\cA:= D_C$, as in \eqref{eq:quadratic-QW-dist-def}.
\ede

\br
    ``Distance'' in the above generalized Wasserstein sense does not necessarily mean a genuine metric. It can produce self-distance and for general cost operators it is not even symmetric in its variables as we shall see. Nevertheless, ``Wasserstein distance'' is the customary terminology in the symmetric special case.
\er

The cost of any $\Pi\in \cS\rbr{\hoh}$ with respect to any positive semidefinite cost operator $C\in\bo{\hoh}$ can be given by an affine combination of quadratic forms of $C$ by elementary tensors of unit length as shown in the next proposition.

\bp[Affine reduction to elementary cost terms]\label{prop:affine reduction}
Let $C\in\bo{\hoh}$ be positive semidefinite and let
$\Pi\in\cS\rbr{\hoh}$. Then there exist finitely many unit vectors
$x_i,y_i\in\cH$ and real coefficients $\alpha_i\in\bR$ such that
\be
    \sum_i \alpha_i = 1
\ee
and
\be\label{eq:cost-affine-form}
    \Cost_C\rbr{\Pi}=
    \sum_i
    \alpha_i
    \bbra{\ketbra{x_i}{y_i}}
    C
    \kket{\ketbra{x_i}{y_i}}.
\ee

Moreover, if $\Pi$ is separable, then the coefficients can be chosen
nonnegative. In that case the above affine combination is a genuine
convex combination. In particular for products of pure states,
\be
    \Cost_C\rbr{\ketbra{\varphi}{\varphi}\otimes\rbr{\ketbra{\psi}{\psi}}^T}=
    \bbra{\ketbra{\varphi}{\psi}}
    C
    \kket{\ketbra{\varphi}{\psi}}.
\ee
\ep

\bpr
By \Cref{prop:doubleket-categories}, write
\be
    \Pi
    =
    \sum_{i} \alpha_{i}\kket{\ketbra{x_i}{y_i}}\bbra{\ketbra{x_i}{y_i}}
\ee
with unit vectors $x_i,y_i\in\cH$ and real coefficients
$\alpha_i\in\bR$. Taking traces gives
\be
    1
    =
    \Tr\Pi
    =
    \sum_i \alpha_i.
\ee

Pairing the above expansion with $C$ gives
\be
    \Cost_C\rbr{\Pi}=
    \sum_{i} \alpha_{i}\Tr C\kket{\ketbra{x_i}{y_i}}\bbra{\ketbra{x_i}{y_i}}=
    \sum_{i} \alpha_{i} \bbra{\ketbra{x_i}{y_i}}C\kket{\ketbra{x_i}{y_i}}.
\ee
Since $C\geq 0$, each elementary cost term appearing on the
right-hand side is nonnegative.

Finally, if $\Pi$ is separable, then by definition it admits a
decomposition
\be
    \Pi
    =
    \sum_{i} p_{i}\kket{\ketbra{x_i}{y_i}}\bbra{\ketbra{x_i}{y_i}},
    \qquad
    p_i\geq 0,
    \qquad
    \sum_i p_i=1,
\ee
and the same computation gives the convex version of the formula.
\epr

Equivalently, the squared distance can be computed from signed decompositions into pure product states.

\bc[Signed affine representation]\label{cor:affine-roof}
Let $C\in\bo{\hoh}$ be positive semidefinite and let
$\rho,\omega\in\cS\rbr{\cH}$. Then 
\be
    D^2_C\rbr{\rho,\omega}&=\inf_{\alpha_i,\psi_i,\varphi_i}\rbr{\sum_i\alpha_iD^2_C\rbr{\ketbra{\psi_i}{\psi_i},\ketbra{\varphi_i}{\varphi_i}}\middle|\sum_i\alpha_i \ketbra{\varphi_i}{\varphi_i}\otimes\rbr{\ketbra{\psi_i}{\psi_i}}^T\in \cC\rbr{\rho,\omega}}\\
    &=\inf_{\alpha_i,\psi_i,\varphi_i}\rbr{\sum_i\alpha_{i} \bbra{\ketbra{\varphi_i}{\psi_i}}C\kket{\ketbra{\varphi_i}{\psi_i}}\middle|\sum_i\alpha_i \ketbra{\varphi_i}{\varphi_i}\otimes\rbr{\ketbra{\psi_i}{\psi_i}}^T\in \cC\rbr{\rho,\omega}},
\ee
where the infimum is over finite families of pairs of vectors $\psi_i,\varphi_i$
    and real coefficients $\alpha_i$. The coupling constraint implies
    $\sum_i\alpha_i=1$.
    If either marginal is pure, \Cref{rem:pure-couplings} gives the unique
    coupling $\ketbra{\varphi_i}{\varphi_i}\otimes \rbr{\ketbra{\psi_i}{\psi_i}}^T$.
\ec

\br
    Joint convexity of $D^2_C$ is immediate from \Cref{cor:affine-roof}. The real linear combination of pure product couplings in \Cref{cor:affine-roof} must itself be positive and have the prescribed marginals;
    these constraints allow entangled couplings. Restricting instead to
    nonnegative coefficients is equivalent to restricting the optimization
    to separable couplings, yielding a convex roof formula, as in
    \cite{tothPitrikQuantumWassersteinSeparable2023}.
\er

There is a bijection between positive semidefinite operators $C\in\bo{\hoh}$ and possible Wasserstein distances defined in \Cref{def:QW-distances} as shown in the following proposition.

\bp\label{prop:diff-cost-diff-dist}
    Let $C_1,C_2\in \bo{\hoh}$ be arbitrary positive semidefinite operators. The following are equivalent.
    \bi
        \ii\label{item:CbijD1} $D_{C_1}\rbr{\rho,\omega}=D_{C_2}\rbr{\rho,\omega}$, for all $\rho,\omega\in \cS\rbr{\cH}$.
        \ii\label{item:CbijD2} $D_{C_1}\rbr{\rho,\omega}=D_{C_2}\rbr{\rho,\omega}$, for all pure states $\rho,\omega\in \cS\rbr{\cH}$.
        \ii\label{item:CbijD3} $C_1=C_2$.
    \ei
\ep
\bpr
    \ref{item:CbijD1} $\Rightarrow$ \ref{item:CbijD2} and \ref{item:CbijD3} $\Rightarrow$ \ref{item:CbijD1} are immediate (by \Cref{cor:affine-roof} \ref{item:CbijD2} $\Rightarrow$ \ref{item:CbijD1} is also clear). For \ref{item:CbijD2} $\Rightarrow$ \ref{item:CbijD3} note that by \Cref{cor:affine-roof} and \Cref{rem:pure-couplings} in particular, for any pure $\rho=\ketbra{\psi}{\psi}$ and any pure $\omega=\ketbra{\varphi}{\varphi}$, \ref{item:CbijD2} can be reformulated as
    \be
        \bbra{\ketbra{\varphi}{\psi}}C_1\kket{\ketbra{\varphi}{\psi}}=D_{C_1}^2\rbr{\rho,\omega}=D_{C_2}^2\rbr{\rho,\omega}=\bbra{\ketbra{\varphi}{\psi}}C_2\kket{\ketbra{\varphi}{\psi}},\qquad \varphi,\psi\in\cH.
    \ee
    Then $C_1=C_2$ follows from \ref{item:doublket-categories2} in \Cref{prop:doubleket-categories}.
\epr

\br
    One aspect of \Cref{prop:affine reduction}, \Cref{cor:affine-roof} and \Cref{prop:diff-cost-diff-dist} can be summarized as follows.
    For arbitrary positive semidefinite cost operator $C$, the whole geometry of a possible definition of the quantum Wasserstein distance is directly encapsulated by
    \bi
        \ii quadratic forms of $C$ on elementary dyads, $\ketbra{\psi}{\varphi}$,
        \be
            \bbra{\ketbra{\psi}{\varphi}}C\kket{\ketbra{\psi}{\varphi}};
        \ee
        \ii and the set and geometry of couplings one allows extremization on.
    \ei
\er

We now record how the possible quadratic Wasserstein distances behave under reversal of the two variables. 
\bde
    Let
    \be
        \ST: \bo{\hoh}\to \bo{\hoh}
    \ee
    denote the swap-transpose map, that is, the linear extension of
    \be
        X\otimes Y^T\mapsto Y\otimes X^T.
    \ee
\ede
\bp\label{prop:ST-properties}
    For the map $\ST$, one has the following properties.
    \bi
        \ii\label{item:ST1} 
            $\ST$ is a self-adjoint involution.
        
        \ii\label{item:ST2} 
            It reverses the ordered pair of marginals:
            \be
                \Pi\in \cC(\rho,\omega)
                \quad\iff\quad
                \ST(\Pi)\in \cC(\omega,\rho).
            \ee
        \ii\label{item:ST3} On the quadratic generators one has
            \be
                \ST\rbr{\ad_X^*\ad_X}
                =
                \ad_X\ad_X^*.
            \ee
        \ii\label{item:ST4} 
            Thus for a family $\cX=\crbr{X_k}_{k=1}^K$, and with $\cX^*:=\crbr{X_k^*}_{k=1}^K$, the quadratic cost operators satisfy
            \be
                \ST\rbr{C_{\cX}}
                =
                C_{\cX^*},\qquad
                \ST\rbr{\widetilde C_{\cX}}
                =
                \widetilde C_{\cX^*}=\widetilde C_{\cX}.
            \ee
    \ei
\ep
\bpr
    For \ref{item:ST1} see \cite{bunthSwapTransposeCouplings2025}. For \ref{item:ST2} see \cite[Proposition 4]{depalmaQuantumOptimalTransport2021}. For \ref{item:ST3}, the action on the generators is given by
        \be
            \ST\rbr{\ad_X}
            =
            \ST\rbr{X\otimes I^T-I\otimes X^T}
            =
           I\otimes X^T-X\otimes I^T
            =-\ad_X.
        \ee
        Note that $\ST$, much like transpose, reverses products and therefore
            \be
            \ST\rbr{\ad_X^*\ad_X}
            =
            \ST\rbr{\ad_X}\ST\rbr{\ad_X^*}
            =
            \ad_X\ad_X^*.
        \ee
    \ref{item:ST4} follows from \ref{item:ST3} and the definitions of the one- and two-sided quadratic cost operators and the symmetry of the latter under $\cX\mapsto \cX^*$.
\epr

\bc\label{cor:symmetry-in-variables}
    For every positive semidefinite cost operator $C$,
    \be
        D_{\ST(C)}(\rho,\omega)=D_C(\omega,\rho),
        \qquad \rho,\omega\in\cS(\cH).
    \ee
    Consequently, $D_C$ is symmetric in its variables if and only if $\ST(C)=C$.
    In particular,
    \be
        D_{\widetilde C_{\cX}}(\rho,\omega)
        =D_{\widetilde C_{\cX}}(\omega,\rho),
        \qquad
        D_{C_{\cX}}(\omega,\rho)=D_{C_{\cX^*}}(\rho,\omega),
    \ee
    and $D_{C_{\cX}}$ is symmetric if and only if $C_{\cX}=C_{\cX^*}$.
\ec
\bpr
    Swap-transpose is a positive, selfadjoint involution and bijects the
    two coupling sets by \Cref{prop:ST-properties}. Thus
    \be
        D_C^2(\omega,\rho)
        =\inf_{\Pi\in\cC(\rho,\omega)}\Tr\ST(\Pi)C
        =\inf_{\Pi\in\cC(\rho,\omega)}\Tr\Pi\ST(C)
        =D_{\ST(C)}^2(\rho,\omega).
    \ee
    The rest of the statements then follow from \Cref{prop:diff-cost-diff-dist} and
    \ref{item:ST4} of \Cref{prop:ST-properties}.
\epr

\br
    \Cref{prop:ST-properties} and \Cref{cor:symmetry-in-variables} highlight another difference of the two-sided and one-sided cost operators: the corresponding Wasserstein distance is forced to be symmetric in its variables for the two-sided cost operators, whereas Wasserstein distances coming from one-sided quadratic cost operators generally may not be.
\er

\section{Equivalent characterization of quadratic costs via Hilbert–Schmidt frame-type operators}\label{sec:frame-characterization}

In this section we isolate a simple linear-algebraic mechanism behind the quadratic cost operators introduced before. The main point is that quadratic costs of observables are not merely associated with families of observables; they are in bijective correspondence with the positive cone generated by Hilbert–Schmidt rank-one projections on the traceless self-adjoint part of $\bo{\cH}$. If $A_1,\ldots,A_K\in\bsa{\cH}$ form a collection of observables, then by \Cref{prop:generalized-cost-gauge-invariance} the scalar parts of the $A_k$'s are invisible to commutators, and the quadratic cost depends only on the traceless components
\be
    \pi_0(A_k)=A_k-\frac{\Tr(A_k)}{d}I.
\ee
The relevant frame-type operator is therefore
\be
    F_{\cA}
    :=
    \sum_{k=1}^K
    \kketbbra{\pi_0(A_k)}{\pi_0(A_k)},
\ee
acting on $\kket{\cV\rbr{d}}$.

The goal is to make the passage $F_{\cA}\iff C_{\cA}$ explicitly invertible and then use it to describe the transport geometry. The forward map is denoted by $\Sigma$, and the recovery map by $\Theta$. After establishing their bijection, we express coupling costs directly through $F_{\cA}=\Theta(C_{\cA})$, specialize to product couplings and pure-state distances, and prove that the rank of this frame operator is the minimal generator number. Finally, isotropic costs correspond exactly to tight frames, and their full frame rank explains why they require the universal maximal number $d^2-1$ of generators in a minimal realization.

\subsection{The frame--cost correspondence}

\bde[Auxiliary definitions]\label{def:auxiliary-definitions}
    For $\cH\simeq \bC^d$, let
    \be
        \mathfrak F_0
        :=
        \crbr{\sum_{k=1}^K
        \kketbbra{A_k}{A_k}
        : K\in\bZ,\; K\geq 0,\;
        A_k\in\widetilde\cV\rbr{d},\; \forall k\in\sqbr{K}}
        \subseteq
        \bo{\hoh}
    \ee
    be the positive cone generated by Hilbert--Schmidt rank-one operators over the traceless self-adjoint part. Equivalently, $\mathfrak F_0$ consists of the positive semidefinite operators $F\in\bo{\hoh}$ satisfying $F\kket{I}=0$ and preserving the real subspace $\kket{\widetilde\cV(d)}$. For positive $F$, the first condition is also equivalent to Hilbert--Schmidt orthogonality to $\kketbbra{I}{I}$. The invariance of the real self-adjoint subspace is an additional condition, not a consequence of positivity and traceless support alone.
    Let
    \be
        \mathfrak C_0
        :=
        \crbr{\sum_{k=1}^K
        \ad_{A_k}^2
        : K\in\bZ,\; K\geq 0,\;
        A_k\in\widetilde\cV\rbr{d},\; \forall k\in\sqbr{K}}
        \subseteq
        \bo{\hoh}
    \ee
    be the positive cone of quadratic cost operators generated by selfadjoint observables. Note that by \Cref{prop:generalized-cost-gauge-invariance}, passing from observables to their traceless parts loses no information.    
    Let
    \be
        \cR:
        \bo{\hoh}
        \to
        \bo{\hoh}
    \ee
    denote the reshuffling map determined by complex-linear extension of
    \be
        \cR\rbr{A\otimes B^T}
        =
        \kketbbra{A}{B^*},
        \qquad
        A,B\in\bo{\cH}.
    \ee
    Equivalently, in matrix entries with respect to an orthonormal basis,
    \be
        \cR(C)_{ij,kl}=C_{ik,jl},\quad C\in\bo{\hoh}.
    \ee
\ede

The following properties are straightforward to check.
\bp \phantom{1}
    \bi
        \ii $\cR^*=\cR^{-1}=\cR$
        \ii $\cR\rbr{I\otimes I^T}=\kketbbra{I}{I}$
        \ii and thus $\cR\rbr{\kketbbra{I}{I}}=I\otimes I^T$.
    \ei
\ep

\bde[The forward map]\label{def:forward-map}
    The forward map
    \be
        \Sigma:\bo{\hoh}\to\bo{\hoh}
    \ee
    is defined by
    \be
        \Sigma(F)
        :=
        \Tr_{\cH^*}F\otimes I^T
        +
        I\otimes \Tr_{\cH}F
        -
        2\cR\rbr{F}.
    \ee
\ede

The following is easy to check but important to note, hence we compute directly.
\bp\label{prop:Sigma-annihilates-tracial}
    The forward map $\Sigma$ annihilates the tracial component $\kketbbra{I}{I}$.
\ep
\bpr
    \be
        \Sigma\rbr{\kketbbra{I}{I}}
        &=
        \Tr_{\cH^*}\kketbbra{I}{I}\otimes I^T
        +
        I\otimes \Tr_{\cH}\kketbbra{I}{I}
        -
        2\cR\rbr{\kketbbra{I}{I}}\\
        &=I\otimes I^T+I\otimes I^T-2I\otimes I^T=0.
    \ee
\epr
The image of $\mathfrak F_0$ under the forward map is $\mathfrak C_0$.
\bp\label{prop:forward-map-is-good}
    Let $\cA:=\crbr{A_k}_{k=1}^K\subseteq\bo{\cH}$ be a collection of observables, and let
    \be
        F_{\cA}
        :=
        \sum_{k=1}^K
        \kketbbra{\pi_0(A_k)}{\pi_0(A_k)}\in \mathfrak F_0
    \ee
    be the associated Hilbert--Schmidt frame-type operator on the traceless part. Then
    \be
    \Sigma\rbr{F_{\cA}}
    =
    \sum_{k=1}^K
    \rbr{
    A_k\otimes I^T-I\otimes A_k^T
    }^2\in \mathfrak C_0.
    \ee
    Thus $\Sigma$ maps the Hilbert--Schmidt frame-type operator of the traceless parts of a family of observables to its associated quadratic cost operator.
\ep

\bpr
    By \eqref{eq:partial-traces-identity},
    \be
        \Tr_{\cH^*}F_\cA=\sum_{k=1}^K \pi_0\rbr{A_k}^2,\qquad \Tr_{\cH}F_\cA=\sum_{k=1}^K \rbr{\pi_0\rbr{A_k}^2}^T,
    \ee
    whereas $\cR$ acts as
    \be
        \cR\rbr{F_\cA}=\sum_{k=1}^K \pi_0\rbr{A_k}\otimes \pi_0\rbr{A_k}^T.
    \ee
    Therefore
    \be
        \Sigma\rbr{F_\cA}&=\sum_{k=1}^K \pi_0\rbr{A_k}^2\otimes I^T + I\otimes \sum_{k=1}^K \rbr{\pi_0\rbr{A_k}^2}^T - 2\sum_{k=1}^K \pi_0\rbr{A_k}\otimes \pi_0\rbr{A_k}^T\\
        &=\sum_{k=1}^K\rbr{\pi_0\rbr{A_k}\otimes I^T-I\otimes \pi_0\rbr{A_k}^T}^2=\sum_{k=1}^K\rbr{A_k\otimes I^T-I\otimes A_k^T}^2
    \ee
    where in the last step we used \Cref{prop:generalized-cost-gauge-invariance}.
\epr

\bp[The forward map is self-adjoint]
\label{prop:forward-is-self-adjoint}
    The forward map $\Sigma$ is self-adjoint with respect to the Hilbert--Schmidt inner product. 
\ep

\bpr
    The Hilbert--Schmidt adjoints of $\Tr_{\cH^*}$ and $\Tr_{\cH}$ are
    \be
        \rbr{\Tr_{\cH^*}}^*(A)=A\otimes I^T,
        \qquad
        \rbr{\Tr_{\cH}}^*(\tilde A)=I\otimes \tilde A,
        \quad A\in\bo{\cH},\quad \tilde A\in\bo{\cH^*}.
    \ee
    Thus
    \be
        \Sigma
        =
        \rbr{\Tr_{\cH^*}}^*\Tr_{\cH^*}
        +
        \rbr{\Tr_{\cH}}^*\Tr_{\cH}
        -
        2\cR
    \ee
    is self-adjoint, since $\cR^*=\cR$.
\epr

\bde[The recovery map]\label{def:recovery-map}
    The recovery map
    \be
        \Theta:\bo{\hoh}\to\bo{\hoh}
    \ee
    is defined by
    \be
        \Theta\rbr{C}:=&-\frac{1}{2}\cR\rbr{\pi_0\otimes\pi_0^T\rbr{C}}\\
        =&-\frac{1}{2}\cR\rbr{C-\frac{1}{d}\Tr_{\cH^*}C\otimes I^T-\frac{1}{d}I\otimes \Tr_{\cH}C+\frac{\Tr C}{d^2}I\otimes I^T}.
    \ee
\ede
The following is easy to check but important to note, hence we compute directly.
\bp\label{prop:Theta-annihilates-identity}
    The recovery map $\Theta$ annihilates tensor products with identity.
\ep
\bpr
    \be
        \Theta\rbr{A\otimes I^T}=-\frac{1}{2}\cR\rbr{\pi_0\rbr{A}\otimes 0}=0\quad \text{and}\quad \Theta\rbr{I\otimes A^T}=-\frac{1}{2}\cR\rbr{0\otimes \pi_0\rbr{A}^T}=0,\quad A\in \bo{\cH}.
    \ee
\epr
The image of $\mathfrak C_0$ under the recovery map is $\mathfrak F_0$.
\bp\label{prop:recovery-map-is-good}
    Let $\cA:=\crbr{A_k}_{k=1}^K\subseteq\bo{\cH}$ be a collection of observables, and let
    \be
        C_{\cA}
        =
        \sum_{k=1}^K
        \rbr{
        A_k\otimes I^T-I\otimes A_k^T
        }^2\in \mathfrak C_0
    \ee
    be the associated quadratic cost. Then
    \be
        \Theta\rbr{C_{\cA}}
        =
        \sum_{k=1}^K
        \kketbbra{\pi_0(A_k)}{\pi_0(A_k)}\in \mathfrak F_0.
    \ee
    Thus $\Theta$ maps the quadratic cost operator to the associated Hilbert--Schmidt frame-type operator of the traceless parts of the observables.
\ep
\bpr
    \be
        {}&\pi_0\otimes\pi_0^T\rbr{C_{\cA}}=\sum_{k=1}^K \rbr{\pi_0\rbr{A_k^2}\otimes \pi_0\rbr{I}^T+\pi_0\rbr{I}\otimes \pi_0\rbr{A_k^2}^T-2\pi_0\rbr{A_k}\otimes \pi_0\rbr{A_k}^T}\\
        =&-2\sum_{k=1}^K\pi_0\rbr{A_k}\otimes \pi_0\rbr{A_k}^T.
    \ee
    Thus
    \be
        \Theta\rbr{C_{\cA}}
        =
        \cR\rbr{\sum_{k=1}^K\pi_0\rbr{A_k}\otimes \pi_0\rbr{A_k}^T}=\sum_{k=1}^K\kketbbra{\pi_0(A_k)}{\pi_0(A_k)}.
    \ee
\epr

\bt[Bijection between Hilbert--Schmidt frame-type operators and quadratic costs]
\label{thm:bijection-frame-cost}
    Let $\cH\simeq\bC^d$, and let $\mathfrak F_0$ and $\mathfrak C_0$ be the cones introduced in \Cref{def:auxiliary-definitions}. Let $\widehat \Sigma$ and $\widehat \Theta$ be $\Sigma$ and $\Theta$, defined in \Cref{def:forward-map} and \Cref{def:recovery-map}, restricted to $\mathfrak F_0$ and $\mathfrak C_0$ respectively,
    \be
        \widehat\Sigma:\mathfrak F_0\to\mathfrak C_0,
        \qquad
        \widehat\Theta:\mathfrak C_0\to\mathfrak F_0.
    \ee
    Then 
    \be
        \widehat\Theta\circ\widehat\Sigma=\id_{\mathfrak F_0},
        \qquad
        \widehat\Sigma\circ\widehat\Theta=\id_{\mathfrak C_0}.
    \ee
    Consequently, consider any two finite collections of operators
    \be
        \cX=\crbr{X_k}_{k=1}^K=\crbr{X_{kS}+iX_{kA}}_{k=1}^K\subseteq\bo{\cH},
        \qquad
        \cY=\crbr{Y_l}_{l=1}^L=\crbr{Y_{lS}+iY_{lA}}_{l=1}^L\subseteq\bo{\cH},
    \ee
    with
    \be
        X_{kS}:=\frac{X_k+X_k^*}{2},\quad X_{kA}:=\frac{X_k-X_k^*}{2i}
    \ee
    and similarly for $Y_l$. Then one has
    \be
        \widetilde{C}_\cX
        =
        \widetilde{C}_\cY
    \ee
    if and only if
    \be
        &\sum_{k=1}^K\rbr{
            \kketbbra{\pi_0(X_{kS})}{\pi_0(X_{kS})}
            +\kketbbra{\pi_0(X_{kA})}{\pi_0(X_{kA})}
        }\\
        &\qquad=
        \sum_{l=1}^L\rbr{
            \kketbbra{\pi_0(Y_{lS})}{\pi_0(Y_{lS})}
            +\kketbbra{\pi_0(Y_{lA})}{\pi_0(Y_{lA})}
        }.
    \ee
    In particular, if all $X_k,Y_l$ are selfadjoint, then
    \be
        C_\cX=C_\cY\iff \widetilde C_\cX=\widetilde C_\cY\iff
        \sum_{k=1}^K
        \kket{\pi_0(X_{k})}\bbra{\pi_0(X_{k})}
        =
        \sum_{l=1}^L
        \kket{\pi_0(Y_{l})}\bbra{\pi_0(Y_{l})}.
    \ee
\et

\bpr
    \be
        \widehat\Theta\circ\widehat\Sigma=\id_{\mathfrak F_0}
        \quad\text{and}\quad
        \widehat\Sigma\circ\widehat\Theta=\id_{\mathfrak C_0}
    \ee
    follows straightforwardly from \Cref{prop:forward-map-is-good} and \Cref{prop:recovery-map-is-good}.        
    The final equivalence follows by noting the special form \eqref{eq:ymmetric-terms} of two-sided quadratic cost operators and applying $\Theta$ to both sides of the equality of quadratic cost operators, and conversely by applying $\Sigma$ to both sides of the equality of Hilbert--Schmidt frame-type operators.
\epr

\subsection{Transport costs and minimal generators}

The correspondence also expresses the transport problem directly in terms of the recovered frame operator. We first transfer the cost pairing through the forward map, then specialize to product couplings, and finally identify the minimal number of generators.

\bde
    For any $C_\cA\in\mathfrak C_0$, define
    \be\label{eq:frame-squared-sum}
        B\rbr{C_\cA}:=\Tr_{\cH^*}\Theta(C_\cA)=\Tr_{\cH^*}\sum_{k=1}^K
        \kketbbra{\pi_0(A_k)}{\pi_0(A_k)}=\sum_{k=1}^K \sqbr{\pi_0(A_k)}^2.
    \ee
\ede

\bp[Transport costs in terms of the recovered frame]
\label{prop:frame-transport-cost}
    $B$ depends only on the cost operator, and it can also be recovered directly as
    \be\label{eq:B-from-cost}
        B\rbr{C}
        =\frac1d\Tr_{\cH^*}C-\frac{\Tr C}{2d^2}I.
    \ee
    For every coupling $\Pi\in\cC\rbr{\rho,\omega}$,
    \be\label{eq:frame-cost-pairing}
        \Cost_C\rbr{\Pi}
        =\Tr \Theta\rbr{C}\Sigma\rbr{\Pi}
        =\Tr\rbr{\rho+\omega}B\rbr{C}
        -2\Tr \Theta\rbr{C}\cR\rbr{\Pi}.
    \ee
    Consequently,
    \be\label{eq:distance-via-frame}
        D_C^2\rbr{\rho,\omega}
        =\Tr\rbr{\rho+\omega}B\rbr{C}
        -2\sup_{\Pi\in\cC\rbr{\rho,\omega}}
        \Tr \Theta\rbr{C}\cR\rbr{\Pi}.
    \ee
\ep

\bpr
    Set $\widetilde A_k:=\pi_0(A_k)$ for a realizing family $C=C_{\cA}$.
    Expanding the cost and using $\Tr \widetilde A_k=0$ gives
    \be
        C=B(C)\otimes I^T+I\otimes B(C)^T
          -2\sum_k \widetilde A_k\otimes(\widetilde A_k)^T,
    \ee
    whence
    \be
        \Tr_{\cH^*}C=dB(C)+(\Tr B(C))I,
        \qquad \Tr C=2d\Tr B(C).
    \ee
    This proves \eqref{eq:B-from-cost}.
    Since $C=\Sigma(\Theta(C))$ and $\Sigma$ is Hilbert--Schmidt selfadjoint,
    \be
        \Tr C\Pi=\Tr\Theta(C)\Sigma(\Pi).
    \ee
    Substituting
    $\Sigma(\Pi)=\omega\otimes I^T+I\otimes\rho^T-2\cR(\Pi)$
    gives \eqref{eq:frame-cost-pairing}; minimizing gives
    \eqref{eq:distance-via-frame}.
\epr

\bde
    For $F\in\mathfrak F_0$ and $\rho\in\cS\rbr{\cH}$, define
    \be\label{eq:frame-total-variance}
        \Var_F(\rho)
        :=\Tr\rbr{\rho\Tr_{\cH^*}F}-\bbra\rho F\kket\rho.
    \ee
    If $F=\sum_k\kketbbra{A_k}{A_k}$ with $A_k\in\widetilde\cV(d)$, then
    $\Var_F(\rho)=\sum_k\rbr{\Tr\rho A_k^2-(\Tr\rho A_k)^2}$.
\ede

\bc[Product couplings, variances, and pure-state distances]
\label{cor:product-coupling-frame}
    Let $C\in\mathfrak C_0$, $F:=\Theta(C)$. For all states $\rho,\omega\in\cS\rbr{\cH}$,
    \be\label{eq:product-cost-frame-linear}
        \Cost_C\rbr{\omega\otimes\rho^T}
        =\Tr\rho \rbr{\Tr_{\cH^*}F}+\Tr\omega \rbr{\Tr_{\cH^*}F}
        -2\bbra{\rho}F\kket{\omega},
    \ee
    or equivalently,
    \be\label{eq:product-cost-frame-variance}
        \Cost_C\rbr{\omega\otimes\rho^T}
        =\Var_F(\rho)+\Var_F(\omega)
        +\bbra{\rho-\omega}F\kket{\rho-\omega}.
    \ee
    These expressions are upper bounds for $D_C^2\rbr{\rho,\omega}$, and are equal to it whenever either state is pure.
\ec

\bpr
    $B\rbr{C}=\Tr_{\cH^*}F$. Then the reshuffling identity
    $\cR\rbr{\omega\otimes\rho^T}=\kketbbra{\omega}{\rho}$
    and \Cref{prop:frame-transport-cost} give
    \eqref{eq:product-cost-frame-linear} and equivalently \eqref{eq:product-cost-frame-variance}. 
\epr

\br
    By the frame-cost operator bijection developed in this section, the frame operator naturally determines both the fixed marginal contribution to the transport cost and the coupling-dependent correlation term. These connections are now made explicit by \eqref{eq:distance-via-frame},  \eqref{eq:product-cost-frame-linear} and \eqref{eq:product-cost-frame-variance}. For product couplings, the frame operator measures a quadratic separation in addition to the two total variances.
\er

\bde[Minimal generator number]\label{def:minimal-generator-number}
    For $C\in\mathfrak C_0$, define
    \be\label{eq:minimal-generator-number}
        m(C):=\min\crbr{
            K\geq0:
            C=\sum_{k=1}^K\ad_{A_k}^2,
            \quad A_k\in\bsa{\cH}
        },
    \ee
    where $K$ is an integer and the empty sum realizes the zero cost. By scalar-shift invariance, restricting the generators to be traceless gives the same minimum.
\ede

\bc[The universal generator bound]\label{cor:minimal-generator-rank}
    For every $C\in\mathfrak C_0$,
    \be\label{eq:minimal-generator-rank}
        m(C)=\rank\Theta(C)\leq d^2-1.
    \ee
    In fact, $F:=\Theta(C)$ admits a spectral decomposition
    with $\lambda_j>0$ and Hilbert--Schmidt orthonormal
    $H_j\in\widetilde\cV(d)$, giving the minimal realization
    \be\label{eq:minimal-spectral-realization}
        F=\sum_{j=1}^{\rank F}\lambda_j\kketbbra{H_j}{H_j},
        \qquad C=\sum_{j=1}^{\rank F}\ad_{\sqrt{\lambda_j}H_j}^2.
    \ee
\ec
\bpr
    A realization with $K$ selfadjoint generators gives a sum of $K$
    rank-one terms for $F$, so $\rank F\leq K$.
    Since $F$ preserves the real Hilbert space $\kket{\widetilde\cV(d)}$,
    its real spectral theorem supplies the indicated selfadjoint
    eigenvectors. Applying $\Sigma$ gives a realization with $\rank F$
    generators. Finally, $F\kket{I}=0$ implies $\rank F\leq d^2-1$.
\epr

\br
    If arbitrary operators are allowed as two-sided generators, the minimal
    number is exactly $\lceil m(C)/2\rceil$.
    Indeed, each such generator supplies two selfadjoint generators by
    \eqref{eq:ymmetric-terms}. Conversely, pair a minimal selfadjoint
    family as $A_1+iA_2,A_3+iA_4,\ldots$, adding a zero last term when needed.
\er

\br
    The number of members of a chosen family is not intrinsic: one may split a generator into proportional generators with the same total squared coefficient, or add scalar observables, without changing the cost. In contrast, $m(C)$ is intrinsic, and \Cref{cor:minimal-generator-rank} shows that no observable-generated quadratic cost can require more than $d^2-1$ self-adjoint generators.
\er

\br[One-sided costs and the frame maps]
    Write $X_k=X_{k,S}+iX_{k,A}$ with selfadjoint real and imaginary parts,
    and set
    \be
        K_{\cX}:=\sum_k i\sqbr{X_{k,S},X_{k,A}},\qquad
        F_{\cX}^{\mathrm{sa}}:=\sum_k\rbr{
            \kketbbra{\pi_0(X_{k,S})}{\pi_0(X_{k,S})}
            +\kketbbra{\pi_0(X_{k,A})}{\pi_0(X_{k,A})}}.
    \ee
    Equations \eqref{eq:nonsymmetric-terms}--\eqref{eq:ymmetric-terms} give
    $C_{\cX}=\widetilde C_{\cX}+\ad_{K_{\cX}}$.
    By \Cref{prop:Theta-annihilates-identity}, $\Theta$ annihilates tensor products with an identity factor and thus
    \be
        \Theta(C_{\cX})=F_{\cX}^{\mathrm{sa}},
        \qquad
        \Sigma\Theta(C_{\cX})=\widetilde C_{\cX}
        =\frac12\rbr{C_{\cX}+C_{\cX^*}}.
    \ee
    Thus $\Sigma\Theta$ is an idempotent retraction of one-sided costs
    onto $\mathfrak C_0$, discarding the additional commutator term.
    Because $K_{\cX}$ is traceless and selfadjoint,
    \be\label{eq:one-sided-is-two-sided}
        C_{\cX}\in\mathfrak C_0
        \quad\iff\quad
        K_{\cX}=0.
    \ee
    Necessity follows because observable-generated costs preserve
    selfadjointness, whereas $\ad_{K_{\cX}}$ takes selfadjoint operators
    to skew-adjoint ones; sufficiency follows from the decomposition.
    For a single generator, \eqref{eq:one-sided-is-two-sided} holds if and only if $X$ is normal.
\er

\subsection{Isotropic costs and tight frames}
For isotropic cost operators and frames, the correspondence takes a particularly simple form.
\bc\label{cor:symmetric-costs}
In every finite dimension and for every $\alpha\geq0$,
\be
    \qquad
    \alpha P_{\kket{\cV\rbr{d}}}
    \xleftrightarrow{\ \Sigma,\Theta\ }
    2d\alpha P_{\kket{\cV\rbr{d}}}
    \qquad
\ee
is the isotropic frame--cost correspondence.
\ec
\bpr
    By \Cref{prop:Sigma-annihilates-tracial}, $\Sigma$ annihilates $\kketbbra{I}{I}$ and therefore,
    \be
        \Sigma\rbr{\alpha P_{\kket{\cV(d)}}}
        &=\alpha\Sigma(I\otimes I^T)
        =
        \alpha\sqbr{\Tr_{\cH^*}\rbr{I\otimes I^T}\otimes I^T
        +
        I\otimes \Tr_{\cH}\rbr{I\otimes I^T}
        -
        2\cR\rbr{I\otimes I^T}}\\
        &=2\alpha\rbr{dI\otimes I^T-\kketbbra{I}{I}}
        =2d\alpha P_{\kket{\cV(d)}}.
    \ee
    The inverse identity follows from the bijection property established in \Cref{thm:bijection-frame-cost}.
\epr
\bex
For example, in dimension $d=2$, the Pauli matrices satisfy
\be
    \Tr\rbr{\sigma_i\sigma_j}=2\delta_{ij},
    \qquad
    i,j\in\crbr{x,y,z}.
\ee
Thus
\be
    \kketbbra{\sigma_x}{\sigma_x}
    +
    \kketbbra{\sigma_y}{\sigma_y}
    +
    \kketbbra{\sigma_z}{\sigma_z}
    =
    2P_{\kket{\cV\rbr{d}}},
\ee
and the associated full Pauli cost is
\be
    C_{\mathrm{sym}}
    =
    \ad_{\sigma_x}^2
    +
    \ad_{\sigma_y}^2
    +
    \ad_{\sigma_z}^2
    =
    8P_{\kket{\cV\rbr{d}}}.
\ee
This is the $d=2$, $\alpha=2$ instance of the general formula
\be
    \Sigma\rbr{\alpha P_{\kket{\cV\rbr{d}}}}=2d\alpha P_{\kket{\cV\rbr{d}}}.
\ee
\eex

\bc[Isotropy and maximal minimal generator number]
\label{cor:isotropic-minimal-generators}
    Let $d>1$, $\alpha>0$, and
    $C=2d\alpha P_{\kket{\cV(d)}}$. Then
    \be
        \Theta(C)=\alpha P_{\kket{\cV(d)}},
        \qquad m(C)=d^2-1.
    \ee
    A family $\cA=\crbr{A_k}_{k=1}^K\subseteq\bsa{\cH}$ realizes $C$ if and only if its traceless parts form a Hilbert--Schmidt tight frame with frame constant $\alpha$. In particular, every realization has $K\geq d^2-1$. A realization is minimal exactly when $K=d^2-1$; in that case its normalized traceless parts
    \be
        \crbr{\pi_0(A_k)/\sqrt{\alpha}}_{k=1}^{d^2-1}
    \ee
    form a Hilbert--Schmidt orthonormal basis of $\cV(d)$. Conversely, any such scaled orthonormal basis realizes $C$.
\ec

\bpr
    The frame--cost identity is \Cref{cor:symmetric-costs}; its rank and \Cref{cor:minimal-generator-rank} give the minimal generator number. The characterization of arbitrary realizations follows from \Cref{thm:bijection-frame-cost}. Finally, a tight frame with exactly $d^2-1$ members is a scaled orthonormal basis by \Cref{lemma:sum-of-rank-ones}.
\epr

\br
    Thus nonzero isotropic costs attain the universal upper bound on the \emph{minimal} number of self-adjoint generators. This maximality is a consequence of isotropy, not a converse characterization. Indeed, for a Hilbert--Schmidt orthonormal basis $\crbr{H_j}_{j=1}^{d^2-1}$ of traceless self-adjoint operators and positive, unequal coefficients $\lambda_j$, the frame
    \be
        F=\sum_{j=1}^{d^2-1}\lambda_j\kketbbra{H_j}{H_j}
    \ee
    has full rank, so $m(\Sigma(F))=d^2-1$, but $\Sigma(F)$ is not isotropic by \Cref{cor:symmetric-costs}. Isotropy is the additional requirement that all these eigenvalues coincide.
\er

\section{Equivalent characterization of symmetric Wasserstein distances}\label{sec:main-theorem}
We now turn explicitly towards proving the main theorem of the paper.

\bde For a function $D:\cS\rbr{\cH}\times \cS\rbr{\cH}\to \bR$, the $D$-isometry monoid of the state space $\cS\rbr{\cH}$ is defined as 
\be
    \Iso(\cS\rbr{\cH},D)
    =
    \crbr{
    \Phi : \cS\rbr{\cH} \to \cS\rbr{\cH}
    :
    D\bigl(\Phi(\rho),\Phi(\omega)\bigr)
    =
    D(\rho,\omega)
    \ \forall \rho,\omega
    }.
\ee
In particular no injectivity, surjectivity, linearity, or continuity is assumed.
\ede

The following two propositions establish the diametral characterization used in the main theorem. The first also supplies the diameter fact needed in the appendix.

\bp[Diametral pairs for an isotropic cost]\label{prop:diameter-of-S}
    Let $d>1$, $\beta>0$, and
    $C=\beta P_{\kket{\cV(d)}}$. Then
    \be
        \sup_{\rho,\omega\in\cS\rbr{\cH}}D_C\rbr{\rho,\omega}=\sqrt{\beta},
    \ee
    and for all states $\rho,\omega$,
    \be\label{eq:isotropic-diametral-pairs}
        D_C\rbr{\rho,\omega}=\sqrt{\beta}
        \quad\iff\quad
        \rho\omega=0
        \quad\iff\quad
        \Tr\rho\omega=0.
    \ee
    If the states are orthogonal, every coupling in
    $\cC\rbr{\rho,\omega}$ has cost $\beta$, and hence every coupling is optimal.
\ep

\bpr
    The product coupling satisfies
    \be\label{eq:indep-coupling-cost}
        \Cost_C\rbr{\omega\otimes\rho^T}
        =\beta\rbr{1-\frac1d\Tr\rho\omega}.
    \ee
    Consequently,
    \be\label{eq:dist-upper-bound}
        D_C^2\rbr{\rho,\omega}
        \leq\beta\rbr{1-\frac1d\Tr\rho\omega}
        \leq\beta.
    \ee

    Conversely, suppose $\rho\omega=0$, and denote their support projections by $P_\rho$ and $P_\omega$. Every coupling $\Pi\in\cC\rbr{\rho,\omega}$ satisfies
    \be\label{eq:coupling-support-projections}
        \Pi
        =\rbr{P_\omega\otimes P_\rho^T}
        \Pi
        \rbr{P_\omega\otimes P_\rho^T}.
    \ee
    Indeed, the expectation of $(I-P_\omega)\otimes I^T$ in $\Pi$ is zero by the first marginal condition. Positivity then implies that this projection annihilates $\Pi$ on both sides. The second marginal gives the analogous conclusion for $I\otimes(I-P_\rho)^T$, proving \eqref{eq:coupling-support-projections}.

    Orthogonality of the supports and the double-ket product rule give
    \be
        \rbr{P_\omega\otimes P_\rho^T}\kket{I}
        =\kket{P_\omega P_\rho}=0.
    \ee
    Thus 
    \be
        \bbra{I}\Pi\kket{I}=\bbra{I}\rbr{P_\omega\otimes P_\rho^T}\Pi\rbr{P_\omega\otimes P_\rho^T}\kket{I}=\bbra{P_\omega P_\rho}\Pi\kket{P_\omega P_\rho}=0,  
    \ee
    and every coupling has the cost
    \be
        \Tr\Pi C
        =\beta\rbr{1-\frac{1}{d}\bbra{I}\Pi\kket{I}}
        =\beta.
    \ee
    Orthogonal states exist because $d>1$, so the diameter is $\sqrt\beta$. The upper bound \eqref{eq:dist-upper-bound} shows that attaining this diameter forces $\Tr\rho\omega=0$, completing the proof.
\epr

\bp[Pure-state diametral orthogonality forces isotropy]
\label{prop:diametral-orthogonality-rigidity}
    Let $d>1$ and let $C\in\bo{\hoh}$ be nonzero and positive semidefinite,
    with $C\kket{I}=0$. Suppose that every orthogonal pair of pure states
    attains the diameter:
    \be
        PQ=0\quad\Longrightarrow\quad
        D_C(P,Q)=\sup_{\rho,\omega\in\cS(\cH)}D_C(\rho,\omega).
    \ee
    Then, for some $\alpha>0$,
    \be
        C=2d\alpha P_{\kket{\cV(d)}},
        \qquad \Theta(C)=\alpha P_{\kket{\cV(d)}}.
    \ee
\ep
\bpr
    Set $\delta:=\sup_{\rho,\omega}D_C(\rho,\omega)$ and
    $A:=\delta^2 I\otimes I^T-C$. Define the linear map
    $\Gamma_A:\bo{\cH}\to\bo{\cH}$ by
    \be\label{eq:diameter-defect-map}
        \kket{\Gamma_A(X)}:=\cR(A)\kket{X}.
    \ee
    For all pure states $P,Q$, the unique-coupling formula and $\cR^*=\cR$ gives
    \be\label{eq:diameter-defect-pure}
        \Tr Q\Gamma_A(P)
        =\bbra{Q}\cR(A)\kket{P}
        =\Tr\sqbr{A\rbr{P\otimes Q^T}}
        =\delta^2-D_C^2(P,Q)\geq0.
    \ee
    Thus $\Gamma_A(P)\geq0$ for all pure state $P$. By the hypothesis, every vector orthogonal
    to the range of $P$ has zero expectation in $\Gamma_A(P)$, so
    \be\label{eq:Gamma-on-pure}
        \Gamma_A(P)=a(P)P,\qquad a(P)\geq0.
    \ee
    Complete any $P$ to an orthogonal resolution of the identity.
    Linearity gives $\Gamma_A(I)P=a(P)P$. Thus every vector is an
    eigenvector of $\Gamma_A(I)$, so $\Gamma_A(I)=aI$ and $a(P)=a$
    for all $P$.
    Rank-one projections span $\bo{\cH}$, hence
    \be\label{eq:Gamma-is-scalar}
        \Gamma_A=a\,\id_{\bo{\cH}},\qquad
        A=\cR\rbr{aI\otimes I^T}=a\kketbbra{I}{I}.
    \ee
    Consequently $C=\delta^2 I\otimes I^T-a\kketbbra{I}{I}$.
    The condition $C\kket{I}=0$ forces $a=\delta^2/d$, so
    $C=\delta^2P_{\kket{\cV(d)}}$.
    Since $C\neq0$, we have $\delta>0$.
    Taking $\alpha=\delta^2/(2d)$ and applying \Cref{cor:symmetric-costs}
    proves the claim.
\epr

We now state the main theorem of the paper characterizing symmetric Wasserstein distances via equivalent conditions.

\bt\label{thm:main}
    Let $\cH$ be a $d$-dimensional Hilbert space with $d>1$, and let
    $\cA=\crbr{A_k}_{k=1}^K\subseteq\bsa{\cH}$ be a finite family of observables, with $K\geq1$ arbitrary. Let $C_{\cA}$ be its quadratic cost operator and $D_{\cA}$ its Wasserstein distance. The following statements are equivalent.
    \bi
        \ii \label{item:main-theorem-1} The $D_{\cA}$-isometry monoid is precisely the set of Wigner symmetries, that is, the conjugations by unitary or antiunitary operators.

        \ii \label{item:main-theorem-2} $D_{\cA}$ is not identically zero and its restriction to pure states is invariant under unitary conjugations.

        \ii \label{item:main-theorem-3} $C_{\cA}\neq0$ and the cost is invariant under the adjoint unitary representation:
        \be
            \Ad_U^*C_{\cA}\Ad_U
            =
            \rbr{U\otimes\bar U}^*C_{\cA}\rbr{U\otimes\bar U}
            =
            C_{\cA},
            \qquad U\in\bu{\cH}.
        \ee

        \ii \label{item:main-theorem-4} The cost is fully isotropic on the traceless component: there exists $\alpha>0$ such that
        \be
            C_{\cA}
            =
            2d\alpha
            \rbr{
                I\otimes I^T-\frac1d\kketbbra{I}{I}
            }.
        \ee

        \ii \label{item:main-theorem-5} The traceless parts of the given generators form a Hilbert--Schmidt tight frame: there exists $\alpha>0$ such that
        \be
            F_{\cA}
            =
            \sum_{k=1}^K\kketbbra{\pi_0\rbr{A_k}}{\pi_0\rbr{A_k}}
            =
            \alpha
            \rbr{
                I\otimes I^T-\frac1d\kketbbra{I}{I}
            }.
        \ee

        \ii \label{item:main-theorem-6} $m\rbr{C_{\cA}}=d^2-1$ and a family $\cB=\crbr{B_j}_{j=1}^{d^2-1}\subseteq\bsa{\cH}$ realizes the cost, $C_{\cA}=C_{\cB}$, if and only if there exists $\alpha>0$ such that
        \be
            \crbr{
                \frac{1}{\sqrt{\alpha}}\pi_0\rbr{B_j}
            }_{j=1}^{d^2-1}
        \ee
        is a Hilbert--Schmidt orthonormal basis of the complex traceless subspace of $\bo{\cH}$.

        \ii \label{item:main-theorem-7} There exists $\alpha>0$ such that
        \be
            \sup_{\rho,\omega\in\cS\rbr{\cH}}D_{\cA}\rbr{\rho,\omega}
            =\sqrt{2d\alpha},
        \ee
        and for all states $\rho,\omega\in\cS\rbr{\cH}$,
        \be
            D_{\cA}\rbr{\rho,\omega}
            =
            \sup_{\rho',\omega'\in\cS\rbr{\cH}}D_{\cA}\rbr{\rho',\omega'}
            \quad\iff\quad
            \rho\omega=0.
        \ee

        \ii \label{item:main-theorem-8} There exists $\alpha>0$ such that
        \be
            \sup_{\rho,\omega\in\cS\rbr{\cH}}D_{\cA}\rbr{\rho,\omega}
            =\sqrt{2d\alpha},
        \ee
        and for all pure states $P,Q\in P_1\rbr{\cH}$,
        \be
            PQ=0\quad\Longrightarrow\quad D_{\cA}\rbr{P,Q}
            =
            \sup_{\rho,\omega\in\cS\rbr{\cH}}D_{\cA}\rbr{\rho,\omega}.
        \ee
    \ei
    Whenever these conditions hold all $\alpha$ parameters above coincide.
    Furthermore, $D_{\cA}$ is not identically zero if and only if $C_{\cA}\neq0$, which in turn holds if and only if at least one $\pi_0(A_k)$ is nonzero.
\et

\bpr
    \ref{item:main-theorem-1} $\Rightarrow$ \ref{item:main-theorem-2}.
    Every unitary conjugation is a Wigner symmetry. Moreover, if $D_{\cA}$ were identically zero, every map of the state space would be an isometry.

    \ref{item:main-theorem-2} $\Rightarrow$ \ref{item:main-theorem-3}.
    Nonvanishing of $C_{\cA}$ follows from \Cref{prop:diff-cost-diff-dist}. By the uniqueness of pure-state couplings and the double-ket product rule, unitary invariance on pure states gives
    \be
        \bbra{\ketbra{\varphi}{\psi}}
        \rbr{\Ad_U^*C_{\cA}\Ad_U-C_{\cA}}
        \kket{\ketbra{\varphi}{\psi}}=0
    \ee
    for all unit vectors $\varphi,\psi$ and all $U\in\bu{\cH}$.  \Cref{prop:doubleket-categories} therefore gives $\Ad_U^*C_{\cA}\Ad_U=C_{\cA}$ for all $U\in \bu{\cH}$.

    \ref{item:main-theorem-3} $\Rightarrow$ \ref{item:main-theorem-4}.
    By \Cref{prop:irreducibility} and \Cref{cor:decomposition}, the cost acts as a scalar on each of the scalar and traceless components. The scalar component is annihilated by \Cref{cor:cost-maps-to-traceless}. Positivity and nonvanishing then give a strictly positive scalar on the traceless component, which we write as $2d\alpha$.

    \ref{item:main-theorem-4} $\Rightarrow$ \ref{item:main-theorem-1}.
    \Cref{prop:equivalent-characterizations-of-symmetric-cost} shows that every Wigner symmetry is an isometry, and \Cref{prop:isotropy-implies-Wigner} proves the converse. These are the dimension-independent versions of the argument in \cite[Theorem~1]{bunthQuantumWassersteinIsometries2026}, recorded in \Cref{app:recast-proof}.

    \ref{item:main-theorem-4} $\Leftrightarrow$ \ref{item:main-theorem-5}.
    This is the isotropic frame--cost correspondence of \Cref{cor:symmetric-costs}, and it also shows equality of the corresponding $\alpha$ parameters.

    \ref{item:main-theorem-4} $\Leftrightarrow$ \ref{item:main-theorem-6}.
    The forward implication is \Cref{cor:isotropic-minimal-generators}. Conversely, choose a selfadjoint Hilbert--Schmidt orthonormal basis $\crbr{H_j}_{j=1}^{d^2-1}$ of the traceless subspace and apply (vi) to $B_j=\sqrt{\alpha}H_j$. Then \Cref{cor:symmetric-costs} gives (iv), with the same $\alpha$.

    \ref{item:main-theorem-4} $\Rightarrow$ \ref{item:main-theorem-7} $\Rightarrow$ \ref{item:main-theorem-8} $\Rightarrow$ \ref{item:main-theorem-4}.
    The first implication is \Cref{prop:diameter-of-S}, the second is restriction to pure states, and the third is \Cref{prop:diametral-orthogonality-rigidity}, also showing equality of the respective $\alpha$ parameters. 

    Finally, \Cref{prop:diff-cost-diff-dist} and the frame--cost bijection give
    \be
        D_{\cA}\equiv0
        \quad\iff\quad C_{\cA}=0
        \quad\iff\quad F_{\cA}=0
        \quad\iff\quad
        \pi_0(A_k)=0\ \text{for every }k.
    \ee
\epr

\appendix

\section{Recasting of an argument from earlier joint work}
\label{app:recast-proof}

For completeness, we record the dimension-independent version of the argument from our earlier joint work \cite[Theorem~1]{bunthQuantumWassersteinIsometries2026} in a compact form. There it was written for dimensions of the form $2^n$. The same argument applies to the isotropic costs in any finite dimension $d>1$. We use the isotropic diameter characterization proved independently in \Cref{prop:diameter-of-S}; the remaining steps are the preservation of pure states, Wigner's theorem, and determination of a state by its distances to pure states.
We first show that every Wigner symmetry preserves an isotropic cost.

\bp\label{prop:equivalent-characterizations-of-symmetric-cost}
 Let $\cH$ be a Hilbert space with $\dim\rbr{\cH}=d$. Suppose that $\exists\alpha\geq0$, such that $C_\cA=\alpha P_{\kket{\cV(d)}}$. Then the map
\be
    \rho \mapsto U\rho U^*
\ee
is a $D_{\cA}$-isometry of $\cS\rbr{\cH}$ for any unitary or antiunitary $U$.
\ep
\bpr
    For a unitary $U$, the map
    $\Pi\mapsto(U\otimes\bar U)\Pi(U\otimes\bar U)^*$ is a bijection
    \be\label{eq:inclusion-of-couplings}
        \cC(\rho,\omega)\longrightarrow
        \cC(U\rho U^*,U\omega U^*).
    \ee
    Indeed, the partial traces transform by the corresponding unitary
    conjugations, and $U^*$ gives the inverse. Moreover,
    $(U\otimes\bar U)\kket{I}=\kket{I}$, so
    \be\label{eq:unitary-invariance-of-C-sym}
        (U\otimes\bar U)^*C_{\cA}(U\otimes\bar U)=C_{\cA}.
    \ee
    Thus this bijection preserves the cost of every coupling and hence
    the infimum.

    Fix an orthonormal basis and let $J$ and $J'$ be conjugations in this
    basis and its dual basis. The antiunitary $K:=J\otimes J'$ gives the
    analogous coupling bijection $\Pi\mapsto K\Pi K$ for
    $(\rho,\omega)\mapsto(J\rho J,J\omega J)$, and $K\kket{I}=\kket{I}$.
    Since $KC_{\cA}K=C_{\cA}$ and coupling costs are real,
    \be
        \Tr\rbr{K\Pi KC_{\cA}}
        =\Tr\rbr{K\Pi C_{\cA}K}=\overline{\Tr(\Pi C_{\cA})}=\Tr(\Pi C_{\cA}).
    \ee
    Hence conjugation by $J$ is also an isometry. Every antiunitary has
    the form $UJ$ with $U$ unitary, proving the assertion.
\epr

The next step is a metric characterization of pure states.

\bp\label{prop:metric-char-of-pure-states}
   Let $\cH$ be a $d$-dimensional Hilbert space with $d>1$. Suppose that $\exists\alpha>0$, such that $C_\cA=\alpha P_{\kket{\cV(d)}}$. For a state
    $\rho \in \cS\rbr{\cH}$ the following are equivalent:
    \bi
        \ii\label{item:metric-char-first} $\rho \in P_1\rbr{\cH}$, that is, $\rho$ is a pure state.

        \ii\label{item:metric-char-second} There exist states
        \be
            \rho_1,\rho_2,\dots,\rho_{d-1} \in \cS\rbr{\cH}
        \ee
        such that, with $\rho_0=\rho$,
        \be
            D_{\cA}\rbr{\rho_j,\rho_k}
            =
            \sup_{\rho,\omega\in\cS\rbr{\cH}}D_{\cA}\rbr{\rho,\omega}
            =
            \sqrt{\alpha}
        \ee
        for all
        \be
            j,k \in \crbr{0,1,\dots,d-1},
            \qquad
            j\neq k.
        \ee
    \ei
\ep

\bpr
    By \Cref{prop:diameter-of-S}, the pairwise distances are maximal
    exactly when the $d$ supports are pairwise orthogonal.
    A pure state extends to such a family by completing its unit vector
    to an orthonormal basis. Conversely, $d$ nonzero pairwise orthogonal
    subspaces of a $d$-dimensional space must all be one-dimensional,
    so every state in such a family is pure.
\epr

The above characterization of pure states refers only to the symmetric quantum
Wasserstein distance, and hence pure states are preserved by isometries, as
stated in the following corollary.

\bc\label{cor:pure-states-are-preserved}
   Let $\cH$ be a $d$-dimensional Hilbert space with $d>1$. Suppose that $\exists\alpha>0$, such that $C_\cA=\alpha P_{\kket{\cV(d)}}$. If
    \be
        \Phi:\cS\rbr{\cH}\to \cS\rbr{\cH}
    \ee
    is an isometry with respect to the quantum Wasserstein distance
    $D_\cA$, then
    \be
        \Phi\rbr{P_1\rbr{\cH}}\subseteq P_1\rbr{\cH}.
    \ee
    That is, $\Phi\rbr{\rho}$ is a pure state whenever $\rho$ is pure.
\ec

\bpr
    An isometry carries every diametral $d$-tuple to another such tuple.
    Apply \Cref{prop:metric-char-of-pure-states} to a tuple containing
    the given pure state.
\epr

For all pure states $P\in P_1\rbr{\cH}$ and states $\omega\in \cS\rbr{\cH}$, the only coupling is the product coupling and therefore, if $C_\cA=\alpha P_{\kket{\cV(d)}}$, then
\be\label{eq:transition-probability-is-encoded}
    D_{\cA}^2(P,\omega)=\alpha\rbr{1-\frac1d\Tr P\omega}.
\ee
Consequently, every $D_{\cA}$-isometry preserves purity of states by \Cref{cor:pure-states-are-preserved} and transition probabilities on pure states by \eqref{eq:transition-probability-is-encoded}.

Here we recall Wigner's famous theorem
\cite{wignerGruppentheorieUndIhre1931}
(see also \cite{bargmannNoteWignersTheorem1964,geherElementaryProofNonbijective2014,lomontWignerUnitarityAntiunitarityTheorem1963})
describing the structure of transition probability preserving transformations
on pure states. Since our work concerns finite-dimensional Hilbert spaces, we
state the finite-dimensional version of Wigner's theorem for simplicity.

\btu[Wigner's theorem, finite-dimensional, non-bijective version]
    Let $\cH$ be a finite-dimensional Hilbert space, and let
    \be
        T:P_1\rbr{\cH}\to P_1\rbr{\cH}
    \ee
    be a map that preserves transition probabilities between pure
    states, that is,
    \be
        \Tr_{\cH}\sqbr{T\rbr{\rho}T\rbr{\omega}}
        =
        \Tr_{\cH}\sqbr{\rho\omega},
        \qquad
        \rho,\omega\in P_1\rbr{\cH}.
    \ee
    Then $T$ is a unitary or antiunitary conjugation, that is, there exists a
    unitary or antiunitary operator $U:\cH\to\cH$ such that
    \be
        T\rbr{\rho}=U\rho U^*,
        \qquad
        \rho\in P_1\rbr{\cH}.
    \ee
\etu

\bp\label{prop:isotropy-implies-Wigner}
 Let $\cH$ be a $d$-dimensional Hilbert space with $d>1$. Suppose that $\exists\alpha>0$, such that $C_\cA=\alpha P_{\kket{\cV(d)}}$. If
    \be
        \Phi:\cS\rbr{\cH}\to \cS\rbr{\cH}
    \ee
    is an isometry with respect to the quantum Wasserstein distance
    $D_\cA$, then there exists a unitary or antiunitary $U$ such that 
    \be
        \Phi\rbr{\rho}
        =
        U\rho U^*,
        \qquad
        \rho\in\cS\rbr{\cH}.
    \ee
    That is, $\Phi\rbr{\cdot}$ is a unitary or antiunitary conjugation.
\ep

\bpr
    By \Cref{cor:pure-states-are-preserved} and \eqref{eq:transition-probability-is-encoded}, Wigner's theorem supplies a unitary
    or antiunitary $U$ such that $\Phi(P)=UPU^*$ for every pure state $P$.
    By \Cref{prop:equivalent-characterizations-of-symmetric-cost},
    \be\label{eq:Phi-tilde-def}
        \widetilde\Phi(\omega):=U^*\Phi(\omega)U
    \ee
    is an isometry fixing every pure state.
    For pure $P$ and arbitrary $\omega\in\cS(\cH)$, \eqref{eq:transition-probability-is-encoded} gives
    \be\label{eq:on-one-hand}
        \alpha\rbr{1-\frac1d\Tr P\omega}
        =D_{\cA}^2(P,\omega)
        =D_{\cA}^2(P,\widetilde\Phi(\omega))
        =\alpha\rbr{1-\frac1d\Tr P\widetilde\Phi(\omega)}.
    \ee
    Since $\alpha>0$, this yields
    $\Tr P\widetilde\Phi(\omega)=\Tr P\omega$ for all rank-one projections
    $P$. Such projections span $\bo{\cH}$, so
    $\widetilde\Phi(\omega)=\omega$ and therefore $\Phi(\omega)=U\omega U^*$.
\epr

\printbibliography[heading=bibintoc]
\end{document}